\documentclass[%
 reprint,
nofootinbib,
 amsmath,amssymb,
 aps,
]{revtex4-1}

\usepackage{graphicx}% Include figure files
\usepackage{dcolumn}% Align table columns on decimal point
\usepackage{bm}% bold math
\usepackage{xfrac}
\usepackage{color}
\usepackage{hyperref}% add hypertext capabilities
\usepackage{mathtools}
\usepackage{mathrsfs}
\usepackage{subcaption}
\usepackage{aas_macros}
\definecolor{internationalkleinblue}{rgb}{0.0, 0.18, 0.65}
\hypersetup{colorlinks=true, urlcolor=internationalkleinblue, linkcolor=internationalkleinblue, citecolor=internationalkleinblue}

\newcommand\Tstrut{\rule{0pt}{3ex}}

\begin{document}

\preprint{APS/123-QED}

\title{Joint cosmology and mass calibration from tSZ cluster counts and cosmic shear}% Force line breaks with \\

\author{Andrina Nicola}
\email{anicola@astro.princeton.edu}
\affiliation{Department of Astrophysical Sciences, Princeton University, Peyton Hall, Princeton NJ 08544, USA}
\author{Jo Dunkley}
\affiliation{Department of Astrophysical Sciences, Princeton University, Peyton Hall, Princeton NJ 08544, USA}
\affiliation{Department of Physics, Princeton University, Princeton, New Jersey 08544, USA}
\author{David N. Spergel}
\affiliation{Department of Astrophysical Sciences, Princeton University, Peyton Hall, Princeton NJ 08544, USA}
\affiliation{Center for Computational Astrophysics, Flatiron Institute, 162 Fifth Avenue, New York NY 10010, USA}

\date{\today}% It is always \today, today,

\begin{abstract}
We present a new method for joint cosmological parameter inference and cluster mass calibration from a combination of weak lensing measurements and the abundance of thermal Sunyaev-Zel'dovich (tSZ) selected galaxy clusters. We combine cluster counts with the spherical harmonic cosmic shear power spectrum and the cross-correlation between cluster overdensity and cosmic shear. These correlations constrain the cluster mass-observable relation. We model the observables using a halo model framework, including their full non-Gaussian covariance. Forecasting constraints on cosmological and mass calibration parameters for a combination of LSST cosmic shear and Simons Observatory tSZ cluster counts, we find competitive constraints for cluster cosmology, with a factor two improvement in the dark energy figure of merit compared to LSST cosmic shear alone. We find most of the mass calibration information will be in the large and intermediate scales of the cross-correlation between cluster overdensity and cosmic shear. Finally, we find broadly comparable constraints to traditional analyses based on calibrating masses using stacked cluster lensing measurements, with the benefit of consistently accounting for the correlations with cosmic shear. 
\end{abstract}

\maketitle

%\tableofcontents

\section{Introduction}\label{sec:intro}

After the immense progress achieved in the last three decades, observational cosmology is about to undergo transformational changes once again. A number of high-precision, wide-field experiments across the electromagnetic spectrum will soon start operations. Examples include the Rubin Observatory Legacy Survey of Space and Time (LSST)\footnote{\url{https://www.lsst.org/}.}, Euclid\footnote{\url{https://www.euclid-ec.org/}.} and the Roman Telescope\footnote{\url{https://roman.gsfc.nasa.gov/}.} in the optical, as well as the Simons Observatory\footnote{\url{https://simonsobservatory.org/}.} (SO) and CMB Stage 4 (S4) in the microwave, which will deliver galaxy samples of unprecedented size as well as high-precision measurements of Cosmic Microwave Background (CMB) anisotropies, respectively. As the data volume of cosmological surveys increases, these experiments will become increasingly dominated by systematic rather than statistical uncertainties, which will require the development of novel analysis methods.

Galaxy clusters constitute the most massive bound objects in the Universe and their abundance as a function of mass is a powerful probe of cosmology, which has the potential to tightly constrain the amplitude of matter fluctuations, $\sigma_{8}$, and the fractional matter density today, $\Omega_{m}$ (see e.g. \cite{Voit:2005, Allen:2011}). However, this exciting cosmological probe has so far received less attention compared to e.g. cosmic shear or galaxy clustering, as it has been limited by systematic uncertainties related to the determination of cluster masses (see e.g. Refs.~\cite{Carlstrom:2002, Voit:2005, Allen:2011} for a discussion). Galaxy clusters can be detected by several different techniques: (i) in the optical by looking for large overdensities in the galaxy distribution, (ii) in the microwave, through their imprint on the observed CMB temperature anisotropies, the  thermal Sunyaev-Zel'dovich (tSZ) effect \cite{Sunyaev:1970}, and finally (iii) in the X-ray through the emission of the hot gas trapped inside these clusters. All of these methods measure an observable that is connected to mass, such as richness $\lambda$, tSZ decrement $Y$ and gas temperature and density $T, \rho$. The uncertainty in the  mass-observable relation is the largest systematic uncertainty in cosmological analyses of galaxy clusters and needs to be calibrated using external data. Weak gravitational lensing is sensitive to all  matter in the Universe and therefore, the lensing signal for galaxies located behind a given cluster can be used to infer cluster halo masses and calibrate the mass-observable relation (e.g. \cite{Allen:2011}). Examples of recent cosmological analyses of galaxy clusters include Refs.~\cite{Hasselfield:2013, Bocquet:2019, Planck:2016XXIV, Zubeldia:2019}, which use CMB data from the Atacama Cosmology Telescope\footnote{\url{https://act.princeton.edu/}.} (ACT), the South Pole Telescope\footnote{\url{https://pole.uchicago.edu/}.} (SPT) and Planck respectively, as well as Refs.~\cite{Mantz:2014, Abbott:2020}, which use X-ray data from Chandra and optical data from the Dark Energy Survey\footnote{\url{https://www.darkenergysurvey.org/}.} (DES), respectively.

In addition, several recent works have investigated joint constraints on cosmology and cluster mass calibration: for example Ref.~\cite{Madhavacheril:2017} forecasted constraints from a joint analysis of CMB S4 cluster abundances and LSST weak lensing, Ref.~\cite{Salcedo:2020} focused on a combination of cluster weak lensing with galaxy clustering and the cross-correlation between cluster and galaxy overdensity and finally Ref.~\cite{Shirasaki:2020} took a different approach: focusing only on power spectra, the authors investigated the potential of multi-wavelength analyses to jointly constrain cosmology and properties of the intracluster medium.

In this work, we focus on the abundance of galaxy clusters detected through the tSZ effect in CMB temperature anisotropy maps. Building on previous work \cite{Oguri:2011, Shirasaki:2015, Krause:2017}, we propose a new method for joint cosmological parameter inference and cluster mass calibration from a combination of weak lensing measurements and tSZ cluster abundances. Specifically, we combine cluster number counts with the spherical harmonic cosmic shear power spectrum and the cross-correlation between cluster overdensity and cosmic shear. We use a halo model  \cite{Ma:2000, Peacock:2000, Seljak:2000, Cooray:2002} framework for modeling the observables and their full non-Gaussian covariance. Using this framework, we forecast constraints on cosmological and mass calibration parameters for a combination of LSST and SO and investigate the different sources of cosmological and astrophysical information. Finally, we compare our results to those obtained with more traditional tSZ mass calibration methods, which are based on stacked measurements of cluster weak lensing (for a summary of the method, the reader is referred to e.g. Ref.~\cite{Madhavacheril:2017}, for examples of stacked weak lensing analyses, see e.g. Refs.~\cite{Medezinski:2018, Miyatake:2019}). Although we focus on forecasting the constraining power of future experiments in this work, the methods presented here are equally applicable to joint analyses of current surveys, such as ACT, SPT and DES.

This paper is organized as follows. In Sec.~\ref{sec:obs}, we present the cosmological observables used in our analysis. Section \ref{sec:theory} outlines the theoretical modeling of the observables within the halo model and in Sec.~\ref{sec:covariance}, we derive expressions for the joint covariance between the probes considered. Sec.~\ref{sec:lsstxso} describes our fiducial assumptions for forecasting joint constraints from LSST and SO and Sec.~\ref{sec:forecasts} describes the forecasting methodology. We present our results in Sec.~\ref{sec:results} and conclude in Sec.~\ref{sec:conclusions}. Implementation details are deferred to the Appendices 

\section{Observables}\label{sec:obs}

In this work, we investigate the potential of joint analyses of tSZ cluster number counts and cosmic shear to simultaneously calibrate cluster masses and constrain cosmological parameters. To this end, we focus on combining cluster number counts $\mathcal{N}_{\mathrm{cl}}$ with cosmic shear power spectra $C_{\ell}^{\gamma \gamma}$ and cross-correlations between cluster overdensity $\delta_{\mathrm{cl}}$ and cosmic shear, $C_{\ell}^{\delta_{\mathrm{cl}} \gamma}$. In the following, we describe these observables in more detail. Unless stated otherwise, all theoretical predictions in this work assume a flat cosmological model, i.e. $\Omega_{k} = 0$.

\subsection{tSZ cluster number counts}

\subsubsection{Cluster detection}\label{subsubsec:obs.tSZ.detection}

The modeling of both the thermal Sunyaev-Zel'dovich signal and cluster detection in this work closely follows Ref.~\cite{Madhavacheril:2017}. We give a brief summary below but refer the reader to Ref.~\cite{Madhavacheril:2017} for more details.

The thermal Sunyaev-Zel'dovich effect is a secondary anisotropy of the CMB  due to inverse Compton scattering of CMB photons with energetic, free electrons in galaxy clusters (for a review of tSZ cosmology, see e.g. \cite{Carlstrom:2002}). The tSZ effect leads to a characteristic spectral distortion of the CMB blackbody spectrum that is proportional to the integrated pressure along a given direction $\boldsymbol{\theta}$, given by (see e.g. \cite{Carlstrom:2002, Planck:2016XXII})
\begin{equation}
\frac{\Delta T}{T_{\mathrm{CMB}}}(\nu, \boldsymbol{\theta}) = f(\nu) \frac{\sigma_{T}}{m_{e}c^{2}}\int \mathrm{d}l \; P_{e}(l, \boldsymbol{\theta}) \equiv f(\nu)y(\boldsymbol{\theta}).
\end{equation}
In this equation, $f(\nu)$ is defined as $f(\nu) = x \coth{\sfrac{x}{2}} - 4$ with $x=\sfrac{h \nu}{k_{B}T_{\mathrm{CMB}}}$, where $T_{\mathrm{CMB}}$ denotes the CMB temperature, $h$ and $k_{B}$ are the Planck and Boltzmann constants, respectively. Furthermore, $m_{e}$ denotes electron mass, $\sigma_{T}$ is the Thompson cross-section, $P_{e}(l, \boldsymbol{\theta})$ denotes the three-dimensional cluster pressure profile and $\mathrm{d}l$ is the line-of-sight distance in direction $\boldsymbol{\theta}$. Finally, we have defined the dimensionless Compton-y parameter $y(\boldsymbol{\theta})$, which determines the amplitude of the tSZ signal. We model $P_{e}(l, \boldsymbol{\theta})$ following Ref.~\cite{Madhavacheril:2017}, adopting the analytic pressure profile from Ref.~\cite{Arnaud:2010} with the parameter values given in Ref.~\cite{Madhavacheril:2017}.

Following Ref.~\cite{Madhavacheril:2017}, we assume that  a matched-filter applied to a CMB map is used to define a cluster. For each detected cluster, we define the spherical aperture tSZ flux as \cite{Alonso:2016}
\begin{equation}
Y_{500} = \frac{4 \pi}{D_{A}^{2}(z)} \int_{0}^{R_{500}} \mathrm{d}^{2}r\; r^{2} \frac{\sigma_{T}}{m_{e}c^{2}} P_{e}(r),
\end{equation} 
where $D_{A}(z)$ denotes the physical angular diameter distance and $R_{500}$ is the radius where the density equals 500 times the critical density of the Universe at the cluster redshift $z$\footnote{We note that $Y_{500}$ is not a directly observable quantity but can be related to any measurement of the integrated Compton-y parameter.}. For a given multi-frequency CMB experiment, the uncertainties in measuring $Y_{500}$, denoted $\sigma_{N}$, are determined by the noise and resolution of the different frequency maps. In order to compute these uncertainties, we again follow Ref.~\cite{Madhavacheril:2017} and refer the reader to that work for further details.

\subsubsection{Mass-observable relation}\label{subsubsec:obs.tSZ.ym}

As the quantity $Y_{500}$ is obtained by integrating the Compton-y parameter over the cluster's extent, it is a measure for the total thermal energy of the cluster. We thus expect $Y_{500}$ to be a measure for the cluster halo mass $M$\footnote{Here $M$ denotes a generic mass definition and we transform between definitions as needed. The procedure chosen to transform between mass definitions is outlined in Appendix \ref{ap:sec:mass-trans}.}. The relation between the mean flux $\bar{Y}_{500}$ and the underlying halo mass $M$ is the main systematic uncertainty in tSZ cluster cosmology. In this work, we follow Refs.~\cite{Planck:2014XX, Alonso:2016, Madhavacheril:2017} and model this relation as 
\begin{equation}
\begin{aligned}
\bar{Y}_{500}(M_{500}, z) = & Y_{*}\left[\frac{M_{500}}{M_{*}}\right]^{\alpha_{Y}} e^{\beta_{Y}\log^{2}{\left(\sfrac{M_{500}}{M_{*}}\right)}}(1+z)^{\gamma_{Y}} \times \\ &E^{\sfrac{2}{3}}(z)\left[\frac{D_{A}(z)}{100 \; \sfrac{\mathrm{Mpc}}{h}}\right]^{-2},
\end{aligned}
\label{eq:Y-M-mean}
\end{equation}
where $M_{500}$ denotes the mass enclosed within the radius where the density equals 500 times the critical density of the Universe at the cluster redshift. The quantities $\alpha_{Y}$ and $\beta_{Y}$ account for the first and second order mass dependence and $\gamma_{Y}$ parameterizes a redshift dependence, additional to that expected from self-similar evolution. Furthermore, $Y_{*}$ and $M_{*}$ are constants and $E(z) = \sfrac{H(z)}{H_{0}}$. The quantities $H(z)$ and $H_{0}$ denote the Hubble parameter and its present day value, respectively. The distribution of true tSZ fluxes is usually assumed to take a log-normal form around their mean $\bar{Y}_{500}$, i.e. (e.g. \cite{Alonso:2016})
\begin{equation}
\begin{aligned}
p(Y_{500}^{\mathrm{true}}|M_{500}, z) = \frac{1}{\sqrt{2\pi}\sigma_{\log Y_{500}}(M, z)} \times \\e^{-\sfrac{(\log{Y_{500}^{\mathrm{true}}}-\log{\bar{Y}_{500}(M_{500}, z)})^{2}}{2\sigma^{2}_{\log Y_{500}}(M, z)}},
\end{aligned}
\end{equation} 
where we have introduced the intrinsic mass- and redshift-dependent scatter $\sigma_{\log Y_{500}}(M, z)$, which we model as \cite{Madhavacheril:2017}
\begin{equation}
\sigma_{\log Y_{500}}(M, z) = \sigma_{\log Y_{0}}\left[\frac{M_{500}}{M_{*}}\right]^{\alpha_{\sigma}}(1+z)^{\gamma_{\sigma}}.
\label{eq:Y-M-scatter}
\end{equation}
In the above equation, $\alpha_{\sigma}$ and $\gamma_{\sigma}$ parametrize the mass- and redshift-dependence of the intrinsic scatter, respectively.

\subsubsection{Cluster number counts}

The probability to observe a galaxy cluster at redshift $z$ with mass $M$, true tSZ amplitude $Y_{500}^{\mathrm{true}}$ and observed tSZ amplitude $Y_{500}^{\mathrm{obs}}$ is given by
\begin{equation}
p(M, z, Y_{500}^{\mathrm{true}}, Y_{500}^{\mathrm{obs}}) = p(Y_{500}^{\mathrm{obs}}) p(Y_{500}^{\mathrm{true}}|Y_{500}^{\mathrm{obs}}) p(M, z|Y_{500}^{\mathrm{true}}).
\label{eq:cluster_prob}
\end{equation}
Using 
\begin{equation}
p(M, z|Y_{500}^{\mathrm{true}}) = \frac{p(M, z)}{p(Y_{500}^{\mathrm{true}})}  p(Y_{500}^{\mathrm{true}}|M, z),
\end{equation}
we can rewrite Eq.~\ref{eq:cluster_prob} as
\begin{widetext}
\begin{equation}
p(M, z, Y_{500}^{\mathrm{true}}, Y_{500}^{\mathrm{obs}}) = \frac{p(Y_{500}^{\mathrm{obs}})}{p(Y_{500}^{\mathrm{true}})} p(Y_{500}^{\mathrm{true}}|Y_{500}^{\mathrm{obs}}) p(M, z) p(Y_{500}^{\mathrm{true}}|M, z).
\label{eq:cluster-prob}
\end{equation}
\end{widetext}
Here $p(M, z)$ denotes the normalized halo mass function (as we are computing the probability to observe a cluster), $p(Y_{500}^{\mathrm{true}}|M, z)$ is the probability that a cluster of $Y_{500}^{\mathrm{true}}$ at redshift $z$ has halo mass $M$ and finally $p(Y_{500}^{\mathrm{obs}}|Y_{500}^{\mathrm{true}}) = \sfrac{p(Y_{500}^{\mathrm{obs}})}{p(Y_{500}^{\mathrm{true}})} p(Y_{500}^{\mathrm{true}}|Y_{500}^{\mathrm{obs}})$ denotes the survey-specific cluster selection function. The selection function quantifies the probability of measuring $Y_{500}^{\mathrm{obs}}$ for a true tSZ flux $Y_{500}^{\mathrm{true}}$ and is determined by the experimental uncertainties discussed in Sec.~\ref{subsubsec:obs.tSZ.detection}.

If we instead set $p(M, z)$ to the unnormalized halo mass function, i.e. $p(M, z) = \sfrac{\mathrm{d}n}{\mathrm{d}M}$, then Eq.~\ref{eq:cluster-prob} gives us the number of detected clusters with $M, z, Y_{500}^{\mathrm{obs}}, Y_{500}^{\mathrm{true}}$. Therefore, the observed number of thermal Sunyaev-Zel'dovich detected galaxy clusters in redshift bin $i$ with $z \in [z_{i, \mathrm{min}}, z_{i, \mathrm{max}}]$ and tSZ signal amplitude bin $\alpha$ with $Y_{500}^{\mathrm{obs}} \in [Y_{500, \alpha}^{\mathrm{obs}, \mathrm{min}}, Y_{500, \alpha}^{\mathrm{obs}, \mathrm{max}}]$ becomes
\begin{widetext}
\begin{equation}
\mathcal{N}^{i}_{\mathrm{cl}, \alpha}\coloneqq \mathcal{N}_{\mathrm{cl}}(\Delta Y_{500, \alpha}^{\mathrm{obs}}, \Delta z_{i}) = \Omega_{s} \int_{z_{i, \mathrm{min}}}^{z_{i, \mathrm{max}}} \mathrm{d}z \frac{c}{H(z)} \frac{\mathrm{d}V}{\mathrm{d}\chi} \int \mathrm{d}M \frac{\mathrm{d}n}{\mathrm{d}M} 
\int \mathrm{d}Y_{500}^{\mathrm{true}} \int_{Y_{500, \alpha}^{\mathrm{obs}, \mathrm{min}}}^{Y_{500, \alpha}^{\mathrm{obs}, \mathrm{max}}} \mathrm{d}Y_{500}^{\mathrm{obs}} \; p(Y_{500}^{\mathrm{obs}}|Y_{500}^{\mathrm{true}}) p(Y_{500}^{\mathrm{true}}|M, z),
\end{equation} 
\end{widetext}
where we have integrated over halo mass and $Y_{500}^{\mathrm{true}}$, which are not directly observable.
Here, $\sfrac{\mathrm{d}V}{\mathrm{d}\chi} = \chi^{2}$ denotes the comoving volume element in comoving distance and we have performed the integration over solid angle, which for a survey covering a sky fraction $f_{\mathrm{sky}}$, yields $\Omega_{s} = 4\pi f_{\mathrm{sky}}$.
Defining the integrated survey selection function for $Y_{500}^{\mathrm{obs}}$ bin $\alpha$ as
\begin{equation}
S_{\alpha}(Y_{\mathrm{true}}, M, z) =  \int_{Y_{500, \alpha}^{\mathrm{obs}, \mathrm{min}}}^{Y_{500, \alpha}^{\mathrm{obs}, \mathrm{max}}} \mathrm{d}Y_{500}^{\mathrm{obs}} \; p(Y_{500}^{\mathrm{obs}}|Y_{500}^{\mathrm{true}}),
\end{equation}
we finally obtain
\begin{equation}
\begin{aligned}
\mathcal{N}^{i}_{\mathrm{cl}, \alpha} = \Omega_{s} \int_{z_{i, \mathrm{min}}}^{z_{i, \mathrm{max}}} \mathrm{d}z \frac{c}{H(z)} \frac{\mathrm{d}V}{\mathrm{d}\chi} \int \mathrm{d}M \frac{\mathrm{d}n}{\mathrm{d}M} \times \\ \int \mathrm{d}Y_{500}^{\mathrm{true}}\; p(Y_{500}^{\mathrm{true}}|M, z) S_{\alpha}(Y_{500}^{\mathrm{true}}, M, z).
\label{eq:cluster-counts}
\end{aligned}
\end{equation}
Using the results derived in Sec.~\ref{subsubsec:obs.tSZ.detection}, we can obtain an expression for $S_{\alpha}(Y_{500}^{\mathrm{true}}, M, z)$. Let us assume a detection threshold for clusters given by $q \sigma_{N}(M, z)$, where $\sigma_{N}(M, z)$ denotes the noise in the $Y$ measurement for a cluster of halo mass $M$ at redshift $z$ and $q$ is the detection level\footnote{In this work, we set $q=5$, which corresponds to a $5\sigma$ detection threshold and is typical for CMB tSZ detections.}. This leads to \cite{Alonso:2016}
\begin{equation}
\begin{aligned}
S_{\alpha}(Y_{500}^{\mathrm{true}}, M, z) = \int_{\mathrm{max}(q \sigma_{N}, Y_{500, \alpha}^{\mathrm{obs}, \mathrm{min}})}^{Y_{500, \alpha}^{\mathrm{obs}, \mathrm{max}}} \mathrm{d}Y_{500}^{\mathrm{obs}}\; p(Y_{500}^{\mathrm{obs}}|Y_{500}^{\mathrm{true}}).
\end{aligned}
\end{equation}
Assuming a Gaussian distribution for $p(Y_{500}^{\mathrm{obs}}|Y_{500}^{\mathrm{true}})$ given by \cite{Alonso:2016}
\begin{equation}
p(Y_{500}^{\mathrm{obs}}|Y_{500}^{\mathrm{true}}) = \frac{1}{\sqrt{2\pi}\sigma_{N}(M, z)}e^{-\sfrac{(Y_{500}^{\mathrm{obs}} - Y_{500}^{\mathrm{true}})^{2}}{2 \sigma^{2}_N(M, z)}},
\end{equation}
we finally arrive at \cite{Alonso:2016}
\begin{equation}
\begin{aligned}
S_{\alpha}(Y_{500}^{\mathrm{true}}, M, z)  = \frac{1}{2}\left[\mathrm{erf} \left(\frac{Y_{500, \alpha}^{\mathrm{obs}, \mathrm{max}} - Y_{500}^{\mathrm{true}}}{\sqrt{2}\sigma_{N}(M, z)}\right)\right. - \\
\left.\mathrm{erf} \left(\frac{\mathrm{max}(q \sigma_{N}, Y_{500, \alpha}^{\mathrm{obs}, \mathrm{min}}) - Y_{500}^{\mathrm{true}}}{\sqrt{2}\sigma_{N}(M, z)}\right)\right],
\end{aligned}
\end{equation}
where $\sigma_{N}(M, z)$ is fully determined by experimental uncertainties.

\subsection{Power spectra}\label{subsec:obs.ps}

We combine cluster number counts with two different power spectra: the cosmic shear power spectrum and the cross-power spectrum between cluster overdensity and cosmic shear. 

Let us consider two tracers $a, b \in [\gamma_{i}, \delta_{\mathrm{cl}, \alpha}^{j}]$, where $\gamma$ denotes cosmic shear and $\delta_{\mathrm{cl}}$ denotes cluster overdensity. Furthermore $i, j$ label the respective redshift bins and $\alpha$ the tSZ amplitude bin. Employing the Limber approximation \cite{Limber:1953, Kaiser:1992, Kaiser:1998}, we can write their spherical harmonic power spectrum as
\begin{equation}
\begin{aligned}
C_{\ell}^{ab}=\int \mathrm{d} z \; \frac{c}{H(z)} \; \frac{W^{a}\boldsymbol{\left(}\chi(z)\boldsymbol{\right)}W^{b}\boldsymbol{\left(}\chi(z)\boldsymbol{\right)}}{\chi^{2}(z)} \times \\
P_{ab}\left(k=\frac{\ell+\sfrac{1}{2}}{\chi(z)}, z\right),
\end{aligned}
\end{equation}
where $c$ is the speed of light, $\chi(z)$ is the comoving distance and $P_{ab}(k, z)$ denotes the three-dimensional power spectrum between probes $a$ and $b$. The quantity $W^{a}\boldsymbol{\left(}\chi(z)\boldsymbol{\right)}$ is a probe-specific window function, which we discuss next for cosmic shear and cluster overdensity.

\subsubsection{Cosmic shear power spectrum}

Cosmic shear is sensitive to the integrated matter distribution between source galaxies and the observer, and the cosmic shear kernel $W^{\gamma}\boldsymbol{\left(}\chi(z)\boldsymbol{\right)}$ is given by
\begin{equation}
W^{i}_{\gamma}\boldsymbol{\left(}\chi(z)\boldsymbol{\right)} = \frac{3}{2} \frac{\Omega_{m} H^{2}_{0}}{c^{2}} \frac{\chi(z)}{a} \int_{\chi(z)}^{\chi_{h}} \mathrm{d} z' n^{i}(z') \frac{\chi(z')-\chi(z)}{\chi(z')},
\label{eq:gammawindow}
\end{equation}
where $n^{i}(z)$ denotes the normalized redshift distribution of source galaxies in redshift bin $i$, $\chi_{h}$ is the comoving distance to the horizon and $a$ denotes the scale factor. As cosmic shear is sensitive to all gravitationally interacting matter in the Universe, we further set $P_{\gamma \gamma}(k, z) = P_{mm}(k, z)$, where $P_{mm}(k, z)$ denotes the matter power spectrum.

The observed cosmic shear auto-power spectrum receives an additional contribution due to shape noise from intrinsic galaxy ellipticities. We model the shape noise power spectrum of redshift bin $i$ as $N^{i}_{\gamma \gamma} = \sfrac{\sigma_{\epsilon, i}^{2}}{\bar{n}^{i}_{\mathrm{source}}}$, where $\bar{n}^{i}_{\mathrm{source}}$ denotes the mean angular galaxy number density and $\sigma_{\epsilon, i}$ is the standard deviation of the intrinsic ellipticity in each component.

\subsubsection{Cross-correlation between cluster overdensity and cosmic shear} 

Galaxy clusters are a biased tracer of the matter distribution and their clustering properties can therefore be analyzed analogously to galaxy clustering. In this work, we focus on the angular power spectrum between cluster overdensity and cosmic shear, which can be computed by cross-correlating maps of cluster overdensity and galaxy ellipticity. The redshift distribution of galaxy clusters with tSZ amplitudes in $\Delta Y_{500, \alpha}^{\mathrm{obs}}$, detectable by a given survey, is determined by their number density as a function of redshift (see e.g. \cite{Fedeli:2009}). From Eq.~\ref{eq:cluster-counts} we thus obtain
\begin{equation}
\begin{aligned}
\mathcal{N}_{\mathrm{cl}, \alpha}(z)  \coloneqq \mathcal{N}_{\mathrm{cl}}(z, \Delta Y_{500, \alpha}^{\mathrm{obs}}) =  \Omega_{s} \frac{c}{H(z)} \frac{\mathrm{d}V}{\mathrm{d}\chi} \int \mathrm{d}M \frac{\mathrm{d}n}{\mathrm{d}M} \times \\ \int \mathrm{d}Y_{500}^{\mathrm{true}}\; p(Y_{500}^{\mathrm{true}}|M, z) S_{\alpha}(Y_{500}^{\mathrm{true}}, M, z).
\label{eq:nclust_z}
\end{aligned}
\end{equation}
Finally, normalizing Eq.~\ref{eq:nclust_z} to unity by dividing by the total number of observable clusters in tSZ bin $\alpha$ $\mathcal{N}_{\mathrm{cl}, \alpha}=\int \mathrm{d}z \; \mathcal{N}_{\mathrm{cl}, \alpha}(z)$, we obtain the redshift distribution of galaxy clusters as
\begin{equation}
n_{\mathrm{cl}, \alpha}(z) = \frac{\mathcal{N}_{\mathrm{cl}, \alpha}(z)}{\mathcal{N}_{\mathrm{cl}, \alpha}}.
\end{equation}
In addition to considering bins in tSZ amplitude, we can subdivide the galaxy cluster distribution into redshift bins. We denote the resulting distributions by $n^{i}_{\mathrm{cl}, \alpha}(z)$ and the window function $W^{i}_{\delta_{\mathrm{cl}}, \alpha}\boldsymbol{\left(}\chi(z)\boldsymbol{\right)}$ thus becomes
\begin{equation}
W^{i}_{\delta_{\mathrm{cl}}, \alpha}\boldsymbol{\left(}\chi(z)\boldsymbol{\right)} = \frac{H(z)}{c} n^{i}_{\mathrm{cl}, \alpha}(z).
\end{equation}

While the cross-correlation between cosmic shear and cluster overdensity $C_{\ell}^{\gamma \delta_{\mathrm{cl}}}$ is free from observational noise, the auto-correlation of the cluster overdensity $C_{\ell}^{\delta_{\mathrm{cl}}\delta_{\mathrm{cl}}}$ is subject to Poisson noise. In this analysis, we model this noise power spectrum as $N^{i}_{\delta_{\mathrm{cl}, \alpha} \delta_{\mathrm{cl}, \alpha}} = \sfrac{1}{\bar{n}^{i}_{\mathrm{cl}, \alpha}}$, where $\bar{n}^{i}_{\mathrm{cl}, \alpha}$ denotes the mean angular density of galaxy clusters in tSZ amplitude bin $\alpha$ and redshift bin $i$.

\subsubsection{Systematics modeling}\label{subsec:obs.ps.syst}

We account for potential systematic uncertainties in the cosmic shear measurement by including simple models for these systematics in our theoretical predictions\footnote{The main systematic uncertainty for tSZ cluster number counts is the $Y-M$ relation, which we discuss in Sec.~\ref{subsubsec:obs.tSZ.ym}. We note that we do not account for possible halo assembly bias when modeling the cluster overdensity, as the magnitude and significance of the effect are currently a matter of investigation (see e.g. Ref.~\cite{Sunayama:2019}).}. The most important systematics for cosmic shear are photometric redshift uncertainties and multiplicative biases in measured galaxy shapes. 

\paragraph{Photometric redshift uncertainties}
For each tomographic redshift bin $i$, we parameterize the impact of photo-$z$ uncertainties as
\begin{equation}
n_{i}(z) \propto \hat{n}_{i}(z + \Delta z_{i}),
\end{equation}
where $n_{i}$ denotes the true, underlying redshift distribution, while $\hat{n}_{i}$ is estimated from the galaxy photo-$z$s. The parameter $\Delta z_{i}$ allows us to marginalize over potential biases in the mean of the redshift distributions. 

\paragraph{Multiplicative shear bias}
The estimated weak lensing shear $\boldsymbol{\hat{\gamma}}$ is prone to multiplicative calibration uncertainties, which we model as (e.g. \cite{Heymans:2006})
\begin{equation}
\boldsymbol{\hat{\gamma}} = (1 + m_{i}) \boldsymbol{\gamma}.
\end{equation}
In the above equation, $\boldsymbol{\gamma}$ is the true galaxy shear and $m_{i}$ denotes the multiplicative bias parameter for tomographic redshift bin $i$. 

\section{Theoretical modeling}\label{sec:theory}

In this work, we compute nonlinear matter power spectra $P_{mm}(k, z)$ using the \texttt{Halofit} fitting function \cite{Smith:2003} with the revisions by Ref.~\cite{Takahashi:2012}\footnote{This choice is motivated by the fact that the halo model described below is not able to accurately model power spectra in the transition regime between the 1- and 2-halo term \cite{Mead:2015}.}. We compute theoretical predictions for all other three-dimensional power spectra $P_{ab}(k, z)$ using the halo model \cite{Ma:2000, Peacock:2000, Seljak:2000, Cooray:2002}. In this model, the power spectrum is split into two distinct terms, the 1-halo and the 2-halo term. The 1-halo term quantifies clustering within a single halo, while the 2-halo term accounts for the contributions to $P_{ab}(k, z)$ coming from the relative clustering of tracers in different halos. These two quantities can be written as
\begin{equation}
\begin{aligned}
P^{1h}_{ab}(k, z) &= I^{0}_{ab}(k, k, z), \\
P^{2h}_{ab}(k, z) &= I^{1}_{a}(k, z) I^{1}_{b}(k, z) P_{\mathrm{lin}}(k, z), 
\label{eq:pk_1h_2h}
\end{aligned}
\end{equation}
and the total power spectrum then becomes
\begin{equation}
P_{ab}(k, z) = P^{1h}_{ab}(k, z) + P^{2h}_{ab}(k, z).
\label{eq:pk_hm}
\end{equation}
In Equations \ref{eq:pk_1h_2h} and \ref{eq:pk_hm} we have used the general notation (see e.g. \cite{Cooray:2001, Krause:2017})
\begin{equation}
I^{n}_{a_1\cdots a_m}(k_1,\cdots,k_m) = \int \mathrm{d}M \frac{\mathrm{d}n}{\mathrm{d}M}b_{h, n}(M) \left\langle\prod_{i=1}^m \left[\tilde{u}_{a_{i}}(k_{i}, M) \right]\right\rangle, 
\label{eq:halo-mod-intg}
\end{equation}
where $b_{h, n}(M)$ is the $n$-th order halo bias and we define $b_{h,1}(M)\equiv b_{h}(M)$, $b_{h,0}\equiv1$. The quantity $\tilde{u}_{a_{i}}(k_{i}, M)$ is the Fourier transform of the normalized profile of the distribution of a given tracer within a halo of mass $M$ and $\langle \cdots \rangle$ denotes an ensemble average.

In order to model $P_{ab}(k, z)$, we additionally need expressions for the normalized density profiles for all probes considered, which we will discuss next.

\subsection{Cosmic shear}

Cosmic shear is sensitive to all matter in the Universe and we can therefore employ the halo model quantities for the matter distribution when predicting the statistical properties of cosmic shear. We define $\tilde{u}_{m}(k, M) \equiv \sfrac{M}{\bar{\rho_{m}}} \; u_{m}(k, M)$, where $\bar{\rho}_{m}$ denotes the comoving matter density, and set $\tilde{u}_{\gamma}(k, M) = \tilde{u}_{m}(k, M)$. We further assume a Navarro-Frenk-White profile \cite{Navarro:1996} for the Fourier transform of the matter distribution inside a halo of mass $M$, i.e. \cite{Navarro:1996}
\begin{widetext}
\begin{equation}
u_{m}(k, M) = \left[{\rm ln}(1+c)-\frac{c}{1+c}\right]^{-1}  \left\{\sin x\left[{\rm Si}\left((1+c)\,x\right)-{\rm Si}(x)\right]+ \cos x\left[{\rm Ci}\left((1+c)x\right)-{\rm Ci}(x)\right]-\frac{\sin(cx)}{(1+c)x}\right\},
\end{equation}
\end{widetext}
where $x=\sfrac{k R_\Delta}{c}$, $R_\Delta$ denotes the halo radius, $c=c(M)$ is the concentration parameter, and ${\rm Si}/{\rm Ci}$ denote the sine and cosine integral functions.

\subsection{Galaxy cluster overdensity}

We follow Refs.~\cite{Huetsi:2008, Krause:2017} and assume that each halo of mass $M$ contains at most one galaxy cluster, which is located at its center. In order to derive the Fourier transform of the normalized cluster density profile, we first consider the number density of galaxy clusters in redshift bin $i$ and tSZ amplitude bin $\alpha$ as a function of position $\mathbf{r}$. This can be written as
\begin{equation}
\begin{aligned}
n^{i}_{\mathrm{cl}, \alpha}(\mathbf{r}) = \sum_{\substack{z \in \Delta z_{i},\\ j}} \int \mathrm{d}Y_{500}^{\mathrm{true}}\; p(Y_{500}^{\mathrm{true}}|M, z) \times \\ S_{\alpha}(Y_{500}^{\mathrm{true}}, M, z) \delta_{\mathcal{D}}(\mathbf{r}_{j}),
\end{aligned}
\end{equation}
where $\delta_{\mathcal{D}}(\mathbf{r})$ denotes the Dirac delta function. Switching from discrete to continuous variables, we obtain the mean cluster density in the tSZ and redshift bin as
\begin{equation}
\bar{n}^{i}_{\mathrm{cl}, \alpha} = \int \mathrm{d}M \frac{\mathrm{d}n}{\mathrm{d}M} \int \mathrm{d}Y_{500}^{\mathrm{true}}\; p(Y_{500}^{\mathrm{true}}|M, z) S_{\alpha}(Y_{500}^{\mathrm{true}}, M, z).
\end{equation}
Finally, using the fact that the Fourier transform of the Dirac delta function equals unity, we obtain 
\begin{equation}
\tilde{u}^{i}_{\delta_{\mathrm{cl}, \alpha}}(k, M) = \frac{\int \mathrm{d}Y_{500}^{\mathrm{true}}\; p(Y_{500}^{\mathrm{true}}|M, z) S_{\alpha}(Y_{500}^{\mathrm{true}}, M, z)}{\bar{n}^{i}_{\mathrm{cl}, \alpha}}.
\end{equation}

\subsection{Halo model implementation}

We compute the halo mass function $\sfrac{\mathrm{d}n}{\mathrm{d}M}$ and the halo bias $b_{h}(M)$ using the fitting functions derived in Ref.~\cite{Sheth:1999}. We further assume the concentration-mass relation of halos $c(M)$ to follow the fitting function derived in Ref.~\cite{Duffy:2008}. Unless noted otherwise (e.g. $M_{500}$), halo masses are defined with respect to the mean matter density $\bar{\rho}_{m}$ and we assume a virial collapse density contrast as given by Ref.~\cite{Bryan:1998}\footnote{We note that we transform $\Delta_{c}$ as given in Ref.~\cite{Bryan:1998} to be relative to the matter density instead of the critical density.}.

We further note that the 2-halo term for matter converges to $P_{\mathrm{lin}}(k, z)$ as $k \rightarrow 0$. This imposes a nontrivial constraint on $I^{1}_{m}(k, z)$ as
\begin{equation}
\int \mathrm{d}M \frac{\mathrm{d}n}{\mathrm{d}M}b_{h}(M) \frac{M}{\bar{\rho}_{m}} = 1. 
\end{equation}
We enforce this constraint by adding a constant, correcting for the finite minimal mass cutoff in our halo model integrals. This correction is not necessary for other tracers considered in this work, as these have a physical minimal mass cutoff in all halo model integrals.

In this work, we compute theoretical predictions for cosmological observables using the LSST Dark Energy Science Collaboration (DESC) Core Cosmology Library (CCL\footnote{\url{https://github.com/LSSTDESC/CCL}.}) \cite{Chisari:2019}.

\section{Covariance matrix}\label{sec:covariance}

We compute the joint covariance matrix of cosmic shear, tSZ cluster number counts and the cross-correlation between cosmic shear and cluster overdensity analytically using the halo model. The resulting expressions for all possible combinations between these probes are discussed below. With the exception of the Gaussian covariance of angular power spectra, which does not include mode-coupling effects due to observing only a fraction of the sky (see e.g. Ref.~\cite{Garcia:2019}), these expressions will be useful for both forecasts as well as analyses using real data.

\subsection{Cluster number counts}

The auto-covariance of cluster number counts in redshift bins $i, j$ and tSZ $Y$ bins $\alpha, \beta$ can be subdivided into a Poissonian and a super-sample covariance (SSC) part, i.e.
\begin{equation}
\mathrm{Cov}(\mathcal{N}^{i}_{\mathrm{cl}, \alpha} , \mathcal{N}^{j}_{\mathrm{cl}, \beta}) = \mathrm{Cov}_{\mathrm{P}}(\mathcal{N}^{i}_{\mathrm{cl}, \alpha} , \mathcal{N}^{j}_{\mathrm{cl}, \beta}) +  \mathrm{Cov}_{\mathrm{SSC}}(\mathcal{N}^{i}_{\mathrm{cl}, \alpha} , \mathcal{N}^{j}_{\mathrm{cl}, \beta}).
\end{equation} 
The Poissonian part of the total covariance accounts for the fact that clusters are discrete tracers. The SSC on the other hand quantifies correlations between cluster number counts in different $Y$ bins caused by the presence of long, unresolvable wavelength modes, larger than the survey volume (see e.g. Refs.~\cite{Hamilton:2006, Takada:2013}). 

In this work, we follow Refs.~\cite{Schaan:2014, Krause:2017} and estimate the Poissonian contribution to the total covariance as
\begin{equation}
\mathrm{Cov}_{\mathrm{P}}(\mathcal{N}^{i}_{\mathrm{cl}, \alpha} , \mathcal{N}^{j}_{\mathrm{cl}, \beta}) = \delta^{\mathcal{D}}_{\alpha \beta} \delta^{\mathcal{D}}_{ij} \;\mathcal{N}^{i}_{\mathrm{cl}, \alpha} ,
\end{equation}
where we assume non-overlapping cluster number count bins in tSZ amplitude and redshift and set cross-correlations between cluster number counts at different redshifts to zero.

The super-sample covariance can be estimated as \cite{Takada:2014, Schaan:2014, Krause:2017}
\begin{widetext}
\begin{equation}
\begin{aligned}
\mathrm{Cov}_{\mathrm{SSC}}&(\mathcal{N}^{i}_{\mathrm{cl}, \alpha} , \mathcal{N}^{j}_{\mathrm{cl}, \beta}) = \delta_{ij} \Omega_{s}^{2} \int_{z_{i, \mathrm{min}}}^{z_{i, \mathrm{max}}}  \mathrm{d}z \frac{c}{H(z)} \left[\frac{\mathrm{d}V}{\mathrm{d}\chi}\right]^{2}
\left[\int \mathrm{d}M \frac{\mathrm{d}n}{\mathrm{d}M} b_{h}(M) \int \mathrm{d}Y_{500}^{\mathrm{true}}\; p(Y_{500}^{\mathrm{true}}|M, z) S_{\alpha}(Y_{500}^{\mathrm{true}}, M, z)\right] \times \\ 
&\left[\int \mathrm{d}M' \frac{\mathrm{d}n}{\mathrm{d}M'} b_{h}(M') \int \mathrm{d}Y_{500}^{\prime, \mathrm{true}}\; p(Y_{500}^{\prime, \mathrm{true}}|M', z) S_{\beta}(Y_{500}^{\prime, \mathrm{true}}, M', z)\right] \sigma_b^2(z).
\end{aligned}
\end{equation}
\end{widetext}
The quantity $\sigma_b^2(z)$ is the variance of the long wavelength background mode $\delta_{\rm LS}$ over the survey footprint, given by
 \begin{equation}
\sigma_b^2(z) = \int \frac{\mathrm{d}k_\perp^2}{(2\pi)^2}P_{\rm lin}(k_\perp,z)\left|\tilde{W}(k_\perp,z)\right|^2.
\end{equation}
In the above equation, $\tilde{W}(k_\perp,z)$ denotes the Fourier transform of the survey footprint, which we approximate as a compact circle with an area matched to our data set:
\begin{equation}
\tilde{W}(k_\perp,z)=\frac{2 J_1(k_\perp\chi(z)\theta_s)}{k_\perp \chi(z)\theta_s},\hspace{12pt} \theta_s={\rm arccos}(1-2f_{\rm sky}),
\end{equation}
where $J_1(x)$ is the cylindrical Bessel function of order 1.

\subsection{Angular power spectra}

The covariance of two angular power spectra $C^{ab}_{\ell}$ and $C^{cd}_{\ell'}$ can be written as the sum of a Gaussian, non-Gaussian and super-sample covariance (SSC) part, i.e.
\begin{equation}
\begin{aligned}
\mathrm{Cov}(C^{ab}_{\ell}, C^{cd}_{\ell'}) = &\mathrm{Cov}_{\mathrm{G}}(C^{ab}_{\ell}, C^{cd}_{\ell'}) + \mathrm{Cov}_{\mathrm{NG}}(C^{ab}_{\ell}, C^{cd}_{\ell'}) +\\ &\mathrm{Cov}_{\mathrm{SSC}}(C^{ab}_{\ell}, C^{cd}_{\ell'}). 
\end{aligned}
\end{equation}
The non-Gaussian covariance accounts for mode-coupling due to the non-Gaussianity of the fields being cross-correlated. In analogy to cluster number counts, the SSC quantifies the coupling of small-scale modes due to the presence of long, super-survey modes. 

The Gaussian covariance matrix is given by (see e.g. \cite{Hu:2004, Krause:2017})
\begin{equation}
\begin{aligned}
\mathrm{Cov}_{\mathrm{G}}(C_{\ell}^{ab}, C_{\ell'}^{cd}) = \frac{\delta_{\ell \ell'}}{(2\ell+1)\Delta \ell f_{\mathrm{sky}}} \times \\ 
\left [(C_{\ell}^{ac} + \delta^{\mathcal{D}}_{ac}N^{ac})(C_{\ell}^{bd} + \delta^{\mathcal{D}}_{bd}N^{bd}) \times \right. \\ 
+ \left. (C_{\ell}^{ad} + \delta^{\mathcal{D}}_{ad}N^{ad})(C_{\ell}^{bc} + \delta^{\mathcal{D}}_{bc}N^{bc})\right ],
\label{eq:theorycovmat}
\end{aligned}
\end{equation} 
where $\Delta \ell$ accounts for possible binning of the angular power spectra $C_{\ell}^{ab}$ into bandpowers. The quantities $N^{ab}$ denote the noise power spectra, which are nonzero only for auto-correlations. The expressions for these noise power spectra for the probes considered in our analysis are given in Sec.~\ref{subsec:obs.ps}.

The non-Gaussian covariance is given by the angular projection of the three-dimensional trispectrum\footnote{The trispectrum is the connected part of the four-point function.} $T^{abcd}(k_{1}, k_{2}, k_{3}, k_{4})$ as (see e.g. \cite{Krause:2017})
\begin{widetext}
\begin{equation}
\mathrm{Cov}_{\mathrm{NG}}(C^{ab}_{\ell}, C^{cd}_{\ell'}) = \frac{1}{\Omega_{s}} \int_{\vert \boldsymbol{\ell} \vert \in \ell_{1}} \int_{\vert \boldsymbol{\ell}' \vert \in \ell_{2}} \int \frac{\mathrm{d}^{2}\boldsymbol{\ell}}{A(\ell_{1})} \; \frac{\mathrm{d}^{2}\boldsymbol{\ell}'}{A(\ell_{2})} \; \mathrm{d}\chi \; \frac{W^{a}(\chi)W^{b}(\chi)W^{c}(\chi)W^{d}(\chi)}{\chi^{6}} T^{abcd}(\sfrac{\boldsymbol{\ell}}{\chi}, \sfrac{-\boldsymbol{\ell}}{\chi}, \sfrac{\boldsymbol{\ell}'}{\chi}, \sfrac{-\boldsymbol{\ell}'}{\chi}).
\end{equation}
\end{widetext}
The quantity $A(\ell_{i})$ denotes the area of an annulus of width $\Delta \ell_{i}$ around $\ell_{i}$, i.e. $A(\ell_{i}) \equiv \int_{\vert \boldsymbol{\ell} \vert \in \ell_{i}} \mathrm{d}^{2}\boldsymbol{\ell}$, which is approximately given by $A(\ell_{i}) \approx 2 \pi \Delta \ell_{i} \ell_{i}$ for $\ell_{i} \gg \Delta \ell_{i}$.
    
Using the halo model, the trispectrum $T^{abcd}$ can be written as (e.g. \cite{Takada:2013}):
\begin{equation}
T^{abcd} = T^{abcd, 1h} + (T^{abcd, 2h}_{22} + T^{abcd, 2h}_{13}) + T^{abcd, 3h} + T^{abcd, 4h},
\end{equation}
where
\begin{widetext}
\begin{equation}
\begin{aligned}
T^{abcd, 1h}(\mathbf{k}_{a}, \mathbf{k}_{b}, \mathbf{k}_{c}, \mathbf{k}_{d}) &= I^{0}_{abcd}(k_{a}, k_{b}, k_{c}, k_{d}), \\
T^{abcd, 2h}_{22}(\mathbf{k}_{a}, \mathbf{k}_{b}, \mathbf{k}_{c}, \mathbf{k}_{d}) &= P_{\mathrm{lin}}(k_{ab})I^{1}_{ab}(k_{a}, k_{b})I^{1}_{cd}(k_{c}, k_{d}) + 2 \; \mathrm{perm.}, \\
T^{abcd, 2h}_{13}(\mathbf{k}_{a}, \mathbf{k}_{b}, \mathbf{k}_{c}, \mathbf{k}_{d}) &= P_{\mathrm{lin}}(k_{a})I^{1}_{a}(k_{a})I^{1}_{bcd}(k_{b}, k_{b}, k_{c}) + 3 \; \mathrm{perm.}, \\
T^{abcd, 3h}(\mathbf{k}_{a}, \mathbf{k}_{b}, \mathbf{k}_{c}, \mathbf{k}_{d}) &= B^{\mathrm{PT}}(\mathbf{k}_{a}, \mathbf{k}_{b}, \mathbf{k}_{cd})I^{1}_{a}(k_{a})I^{1}_{b}(k_{b})I^{1}_{cd}(k_{c}, k_{d}) + 5 \;\mathrm{perm.},\\
T^{abcd, 4h}(\mathbf{k}_{a}, \mathbf{k}_{b}, \mathbf{k}_{c}, \mathbf{k}_{d}) &= T^{\mathrm{PT}}(\mathbf{k}_{a}, \mathbf{k}_{b}, \mathbf{k}_{c}, \mathbf{k}_{d})I^{1}_{a}(k_{a})I^{1}_{b}(k_{b})I^{1}_{c}(k_{c})I^{1}_{d}(k_{d}).
\label{eq:halo-mod-trisp}
\end{aligned}
\end{equation}
\end{widetext}
Here, ${\bf k}_{ab}\equiv {\bf k}_a+{\bf k}_b$, and the quantities $B^{\mathrm{PT}}$ and $T^{\mathrm{PT}}$ denote the matter bi- and trispectrum respectively, as estimated using tree-level perturbation theory. The full expressions for these terms can be found in Ref.~\cite{Takada:2013}. For simplicity, we follow \cite{Krause:2017} and approximate the 2- to 4-halo trispectrum as the linearly biased matter trispectrum and only include a probe-specific 1-halo trispectrum contribution. Specifically, we set 
\begin{equation}
T^{abcd} = T^{abcd, 1h} + b_{a}b_{b}b_{c}b_{d}T^{m, 2h+3h+4h},
\end{equation}
where $T^{abcd, 1h}$ and $T^{m, 2h+3h+4h}$ are computed following Equations \ref{eq:halo-mod-trisp}. For $T^{abcd, 1h}$, we evaluate Eq.~\ref{eq:halo-mod-intg} for probes $a, b, d, c$, while for $T^{m, 2h+3h+4h}$, we use the corresponding expressions for the matter distribution. Finally, $b_{a}$ denotes the linear bias of tracer $a$ predicted using the halo model, i.e.
\begin{equation}
b_{a} = \int \mathrm{d}M\,\frac{\mathrm{d}n}{\mathrm{d}M}b_{h}(M) \tilde{u}_{a}(0, M),
\end{equation}
and we set $b_{\gamma}(M) = 1$. From Eq.~\ref{eq:halo-mod-intg}, we see that the 1-halo trispectrum is given by
\begin{widetext}
\begin{equation}
T^{abcd, 1h}(\mathbf{k}_{a}, \mathbf{k}_{b}, \mathbf{k}_{c}, \mathbf{k}_{d}) = \int \mathrm{d}M \frac{\mathrm{d}n}{\mathrm{d}M} 
\left\langle \tilde{u}_{a}(k_{a}, M) \tilde{u}_{b}(k_{b}, M) \tilde{u}_{c}(k_{c}, M) \tilde{u}_{d}(k_{d}, M)\right\rangle.
\end{equation}
\end{widetext}
A special case arises when $T^{abcd, 1h}(\mathbf{k}_{a}, \mathbf{k}_{b}, \mathbf{k}_{c}, \mathbf{k}_{d})$ contains two cluster number count tracers $\delta_{\mathrm{cl}, \alpha}^{i}, \delta_{\mathrm{cl}, \beta}^{j}$ (set to tracers $c, d$ w.l.o.g.), as a halo can at most contain a single cluster. Accounting for this fact, we then obtain
\begin{widetext}
\begin{equation}
T^{abcd, 1h}(\mathbf{k}_{a}, \mathbf{k}_{b}, \mathbf{k}_{c}, \mathbf{k}_{d}) = \delta_{ij}\delta_{\alpha \beta} \int \mathrm{d}M \frac{\mathrm{d}n}{\mathrm{d}M} 
\tilde{u}_{a}(k_{a}, M) \tilde{u}_{b}(k_{b}, M) \frac{\tilde{u}^{i}_{\delta_{\mathrm{cl}, \alpha}}(k_{c}, M)}{(\bar{n}^{i}_{\mathrm{cl}, \alpha})^{2}}.
\end{equation}
\end{widetext}

Finally, we compute the super-sample covariance contribution following the treatment of \cite{Krause:2017}, i.e.:
\begin{widetext}
\begin{equation}
\begin{aligned}
\mathrm{Cov}_{\mathrm{SSC}}(C^{ab}_{\ell}, C^{cd}_{\ell'}) = \int \mathrm{d}\chi \;\frac{W^{a}(\chi)W^{b}(\chi)W^{c}(\chi)W^{d}(\chi)}{\chi^{4}} \frac{\partial P_{ab}(\sfrac{\ell}{\chi}, z(\chi))}{\partial \delta_{\rm LS}}\frac{\partial P_{cd}(\sfrac{\ell'}{\chi}, z(\chi))}{\partial \delta_{\rm LS}}\sigma^{2}_{b}(z(\chi)).
\end{aligned}
\end{equation}
\end{widetext}
The quantity $\partial P_{ab}(k, z)/\partial \delta_{\rm LS}$ denotes the response of the power spectrum $P_{ab}$ to a large-scale density fluctuation, which we estimate using the halo model and results from perturbation theory as (e.g. \cite{Krause:2017}):
\begin{widetext}
\begin{equation}
\begin{aligned}
\frac{\partial P_{ab}(k, z)}{\partial \delta_{\rm LS}} = \left( \frac{68}{21} - \frac{1}{3}\frac{\mathrm{d}\log{k^{3} P_{\mathrm{lin}}}(k, z)}{\mathrm{d}\log k} \right) I_{a}^{1}(k)I_{b}^{1}(k)P_{\mathrm{lin}}(k, z) + I_{ab}^{1}(k, k) - (b_{a, a \neq \gamma} + b_{b, b \neq \gamma})P_{ab}(k, z).
\label{eq:ps-resp}  
\end{aligned}
\end{equation}
\end{widetext}
The last term in Eq.~\ref{eq:ps-resp} accounts for the fact that observed overdensity fields are computed using the mean density estimated inside the survey volume.
    
For consistency with our implementation of the trispectrum, we compute the response function $\sfrac{\partial P_{ab}(k, z)}{\partial \delta_{\rm LS}} $ for a given probe as the linearly biased response of the matter field\footnote{In order to test the robustness of our results to this approximation, we also compute the SSC contribution to the covariance using the probe-specific halo model quantities in Eq.~\ref{eq:ps-resp}. We find our forecasted constraints to be unaffected by this change and therefore resort to the approach described above for consistency.}. 

\subsection{Cross-correlations between cluster number counts and angular power spectra}

Finally, the cross-covariance between cluster number counts and angular power spectra vanishes for purely Gaussian fields, but it receives both non-Gaussian and SSC contributions, i.e.
\begin{equation}
\begin{aligned}
\mathrm{Cov}(\mathcal{N}^{\alpha}_{\mathrm{cl}, i}, C^{ab}_{\ell}) = \mathrm{Cov}_{\mathrm{NG}}(\mathcal{N}^{\alpha}_{\mathrm{cl}, i}, C^{ab}_{\ell}) + \mathrm{Cov}_{\mathrm{SSC}}(\mathcal{N}^{\alpha}_{\mathrm{cl}, i}, C^{ab}_{\ell}). 
\end{aligned}
\end{equation}
Following Refs.~\cite{Takada:2007, Schaan:2014}, we can write the non-Gaussian part of this cross-covariance as
\begin{widetext}
\begin{equation}
\begin{aligned}
\mathrm{Cov}&_{\mathrm{NG}}(\mathcal{N}^{\alpha}_{\mathrm{cl}, i}, C^{ab}_{\ell}) = \Omega_{s} \int_{z_{i, \mathrm{min}}}^{z_{i, \mathrm{max}}} \mathrm{d}z \frac{c}{H(z)} \frac{W^{a}(\chi(z))W^{b}(\chi(z))}{\chi^{4}(z)} \frac{\mathrm{d}V}{\mathrm{d}\chi} \times \\
&\left\{\int \mathrm{d}M \frac{\mathrm{d}n}{\mathrm{d}M} \tilde{u}_{a}(k, M) \tilde{u}_{b}(k, M) \int \mathrm{d}Y_{500}^{\mathrm{true}}\; p(Y_{500}^{\mathrm{true}}|M, z) S_{\alpha}(Y_{500}^{\mathrm{true}}, M, z)+ \right. \\
&\left. \left(\left[\int \mathrm{d}M \frac{\mathrm{d}n}{\mathrm{d}M} b_{h}(M) \tilde{u}_{a}(k, M) \int \mathrm{d}Y_{500}^{\mathrm{true}}\; p(Y_{500}^{\mathrm{true}}|M, z) S_{\alpha}(Y_{500}^{\mathrm{true}}, M, z)\right]
\left[\int \mathrm{d}M \frac{\mathrm{d}n}{\mathrm{d}M} b_{h}(M) \tilde{u}_{b}(k, M)\right] + \right. \right. \\
&\left. \left. \left[\int \mathrm{d}M \frac{\mathrm{d}n}{\mathrm{d}M} b_{h}(M) \tilde{u}_{b}(k, M) \int \mathrm{d}Y_{500}^{\mathrm{true}}\; p(Y_{500}^{\mathrm{true}}|M, z) S_{\alpha}(Y_{500}^{\mathrm{true}}, M, z)\right] 
\left[\int \mathrm{d}M \frac{\mathrm{d}n}{\mathrm{d}M} b_{h}(M) \tilde{u}_{a}(k, M)\right]\right)P_{\mathrm{lin}}(k, z)\right\}.
\end{aligned}
\end{equation}
\end{widetext}
Furthermore, the SSC covariance is given by (see e.g. \cite{Schaan:2014, Krause:2017})
\begin{widetext}
\begin{equation}
\begin{aligned}
\mathrm{Cov}_{\mathrm{SSC}}(\mathcal{N}^{\alpha}_{\mathrm{cl}, i}, C^{ab}_{\ell}) &= \Omega_{s} \int_{z_{i, \mathrm{min}}}^{z_{i, \mathrm{max}}} \mathrm{d}\chi \frac{W^{a}(\chi)W^{b}(\chi)}{\chi^{2}} \frac{\mathrm{d}V}{\mathrm{d}\chi} \left[\int \mathrm{d}M \frac{\mathrm{d}n}{\mathrm{d}M} b_{h}(M) \int \mathrm{d}Y_{500}^{\mathrm{true}}\; p(Y_{500}^{\mathrm{true}}|M, z) S_{\alpha}(Y_{500}^{\mathrm{true}}, M, z)\right] \times \\
&\frac{\partial P_{ab}(\sfrac{\ell}{\chi}, z(\chi))}{\partial \delta_{\rm LS}}\sigma^{2}_{b}(z(\chi)).
\end{aligned}
\end{equation}
\end{widetext}

\section{Combination of LSST and SO}\label{sec:lsstxso}

We assess the potential of a joint analysis of tSZ number counts, cosmic shear and the cross-correlation between cluster overdensity and cosmic shear to simultaneously infer cosmology and mass calibration by performing a Fisher matrix forecast for a combination of LSST and SO\footnote{We note that a similar analysis could be performed for current surveys, such as ACT, SPT and DES.}. The survey specifications assumed for each survey and probe are detailed below.

\subsection{LSST specifications}
We follow Ref.~\cite{Madhavacheril:2017} and model an LSST-like survey assuming a sky coverage of $18'000$ square degrees (corresponding to $f_{\mathrm{sky}}=0.4$), an angular galaxy number density for the weak lensing sample of $\bar{n}_{\mathrm{source}} = 20$ arcmin$^{-2}$ and standard deviation of the intrinsic ellipticity in each component of $\sigma_{\epsilon} = 0.3$. We further assume the redshift distribution of these galaxies to follow the functional form given in Ref.~\cite{Smail:1994}
\begin{equation}
n(z) \propto z^{2} e^{\frac{z}{z_{0}}},
\end{equation}
where we set $z_{0} = 0.3$. The assumed redshift distribution roughly matches the one outlined in the LSST DESC Science Requirements Document \cite{LSST-SRD:2018}, while both the intrinsic ellipticity and angular galaxy number density are more conservative and are derived by extrapolating results from the Hyper-Suprime Cam (HSC) survey \cite{Aihara:2018}. We subdivide the galaxies into four tomographic redshift bins of approximately equal galaxy number between redshift $z_{\mathrm{min}}=0$ and $z_{\mathrm{max}}=3$\footnote{This leads to the following redshift bin edges $z_{\mathrm{min}, i}, z_{\mathrm{max}, i} = [0., 0.57], [0.57, 0.89], [0.89, 1.41], [1.41, 3.]$ for $i = 0, \cdots 3$.} and estimate the true redshift distribution in each photometric redshift bin $i$ using (e.g. \cite{Amara:2007})
\begin{equation}
n_{i}(z_{t}) = \int_{z_{\mathrm{min}, i}}^{z_{\mathrm{max}, i}} \mathrm{d} z_{p} \; p(z_{p}|z_{t}) n(z_{t}),
\end{equation}
where $z_{p}$ denotes photometric and $z_{t}$ true redshift, respectively. Finally, we model $p(z_{p}|z_{t})$ assuming $z_{p}$ to be Gaussian distributed around $z_{t}$ with $\sigma_{p} = 0.05$ \cite{Schaan:2017}.

We compute spherical harmonic power spectra for all auto- and cross-correlations between those redshift bins in 13 angular multipole bins between $\ell_{\mathrm{min}}=100$ and $\ell_{\mathrm{max}}=4600$\footnote{The maximal angular multipole is chosen in accordance with previous LSST forecasts, see e.g. Refs.~\cite{Krause:2017, Schaan:2017}. Furthermore, we choose the bin centers as $\ell_{\mathrm{mean}} = \{100, \allowbreak 200, \allowbreak 300, \allowbreak 400, \allowbreak 600, \allowbreak 800, \allowbreak 1000, \allowbreak 1400, \allowbreak 1800, \allowbreak 2200, \allowbreak 3000, \allowbreak 3800, \allowbreak 4600\}$.}.

\subsection{SO specifications}
We model the expected survey specifications for SO following Ref.~\cite{Ade:2019}, focusing only on the Large Aperture Telescope (LAT). We assume observations in six frequency bandpasses with beam full-width half-maxima (FWHM) and white noise levels for a sky coverage of $f_{\mathrm{sky}}=0.4$ as given in Tab.~\ref{tab:SOspecs} (c.f. Tab. 1 in Ref.~\cite{Ade:2019}). We additionally model the atmospheric noise contribution following Ref.~\cite{Ade:2019} and refer the reader to their Sec. 2.2 for more details. 
\subsubsection{Cluster number counts}
We subdivide the cluster number counts into five bins in redshift between $z_{\mathrm{min}}=0$ and $z_{\mathrm{max}}=1.5$. The maximal cluster redshift is chosen in order to ensure a large enough source sample for mass calibration. Furthermore, photometric redshift uncertainties for LSST are expected to increase significantly at high redshift, which will further limit the usage of high redshift galaxies for mass calibration. We subdivide each of these redshift bins into roughly 15 tSZ amplitude bins between $Y^{\mathrm{obs}}_{500, \mathrm{min}} = 4\times 10^{-13}$ and $Y^{\mathrm{obs}}_{500, \mathrm{max}} = 3\times 10^{-8}$. The exact bin edges and bin numbers considered depend on the cluster redshift bin, as we follow observational analyses (see e.g. \cite{Haan:2016}) and ensure that each bin contains at least a single galaxy cluster\footnote{We note that not applying this cut results in significantly tighter constraints on mass-calibration parameters. However, we choose to not include low cluster number count bins for two reasons: (i) these bins mainly correspond to the high mass end of the mass function, where the approximations made for computing the covariance matrix in this work might break down, and (ii) including bins with very few objects can cause numerical instabilities in Fisher matrix computations.}. The exact bin configurations are given in Appendix \ref{ap:sec:implementation.counts}.

\subsubsection{Cluster lensing}
In order to measure the cluster lensing cross-correlation $C_{\ell}^{\gamma \delta_{\mathrm{cl}}}$, we subdivide the cluster overdensity field into four redshift bins between $z_{\mathrm{min}}=0$ and $z_{\mathrm{max}}=1.41$ and four tSZ amplitude bins between $Y^{\mathrm{obs}}_{500, \mathrm{min}} = 4\times 10^{-13}$ and $Y^{\mathrm{obs}}_{500, \mathrm{max}} = 1.4\times 10^{-9}$. We remove five bins from this subdivision, as they contain less than one cluster, which leaves us with 11 cluster overdensity bins\footnote{The exact bin configurations are given in Appendix \ref{ap:sec:implementation.clust-lens}.}. Furthermore, we only include cross-correlations between galaxy cluster overdensity and cosmic shear for bin combinations for which the lenses are located behind the clusters. These specifications leave us with 20 cross-power spectra $C_{\ell}^{\gamma^{i} \delta^{j}_{\mathrm{cl}, \alpha}}$, which we compute for 16 angular multipole bins between $\ell_{\mathrm{min}}=100$ and $\ell_{\mathrm{max}}=9400$\footnote{This choice of maximal angular multipole ensures that we include a significant amount of information coming from the 1-halo term and is similar to earlier analyses, e.g. \cite{Krause:2017}. Furthermore, the bin centers are chosen as $\ell_{\mathrm{mean}} = \{100, \allowbreak 200, \allowbreak 300, \allowbreak 400, \allowbreak 600, \allowbreak 800, \allowbreak 1000, \allowbreak 1400, \allowbreak 1800, \allowbreak 2200, \allowbreak 3000, \allowbreak 3800, \allowbreak 4600, \allowbreak 6200, \allowbreak 7800, \allowbreak 9400\}$.}. 

\begin{figure*}
\begin{center}
\includegraphics[width=0.98\textwidth]{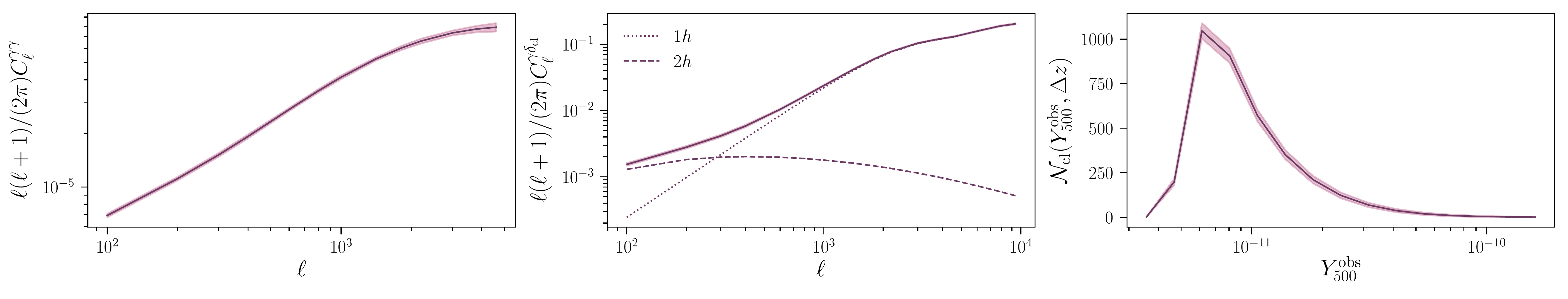}
 \caption{Examples of the observables considered in this analysis. The leftmost panel shows the cosmic shear auto-power spectrum for redshift bin $i=1$ ($z_{\mathrm{min}} = 0.57$, $z_{\mathrm{max}} = 0.89$), the middle panel shows the cross-correlation between cosmic shear bin $i=3$ ($z_{\mathrm{min}} = 1.41$, $z_{\mathrm{max}} = 3.$) and cluster overdensity bin $i=1$, $\alpha=1$ ($z_{\mathrm{min}} = 0.35$, $z_{\mathrm{max}} = 0.7$, $Y_{\mathrm{min}} = 3.08 \times 10^{-12}$, $Y_{\mathrm{max}} = 2.4 \times 10^{-11}$) and finally the last panel shows the cluster number counts for redshift bin $i=2$ ($z_{\mathrm{min}} = 0.5$, $z_{\mathrm{max}} = 0.75$). We have subdivided the cluster lensing power spectrum into its 1-halo and 2-halo contribution. In all panels, the shaded regions show the 1 $\sigma$ uncertainties.}
\label{fig:observables}
\end{center}
\end{figure*}

Finally, when combining LSST and SO, we assume full overlap between the two surveys over a fraction of the sky $f_{\mathrm{sky}}=0.4$. Fig.~\ref{fig:observables} shows an example for each of the three observables considered in our analysis, computed according to the survey and binning specifications given above. 

\begin{table}
\caption{Summary of assumed survey specifications for SO LAT (see also Tab. 1 in Ref.~\cite{Ade:2019}).} \label{tab:SOspecs}
\begin{center}
\begin{ruledtabular}
\begin{tabular}{ccc}
Frequency [GHz] & FWHM [arcmin] & Noise (goal) [$\mu$K arcmin]  \\ \hline \Tstrut           
27 & 7.4 & 52  \\
39 & 5.1 & 27  \\
93 & 2.2 & 5.8  \\
145 & 1.4 & 6.3  \\
225 & 1.0 & 15  \\
280 & 0.9 & 37 
\end{tabular}
\end{ruledtabular}
\end{center}
\end{table} 

\section{Methodology for joint cosmology and mass calibration}\label{sec:mass-calib}

\begin{figure*}
\begin{center}
\includegraphics[width=0.95\textwidth]{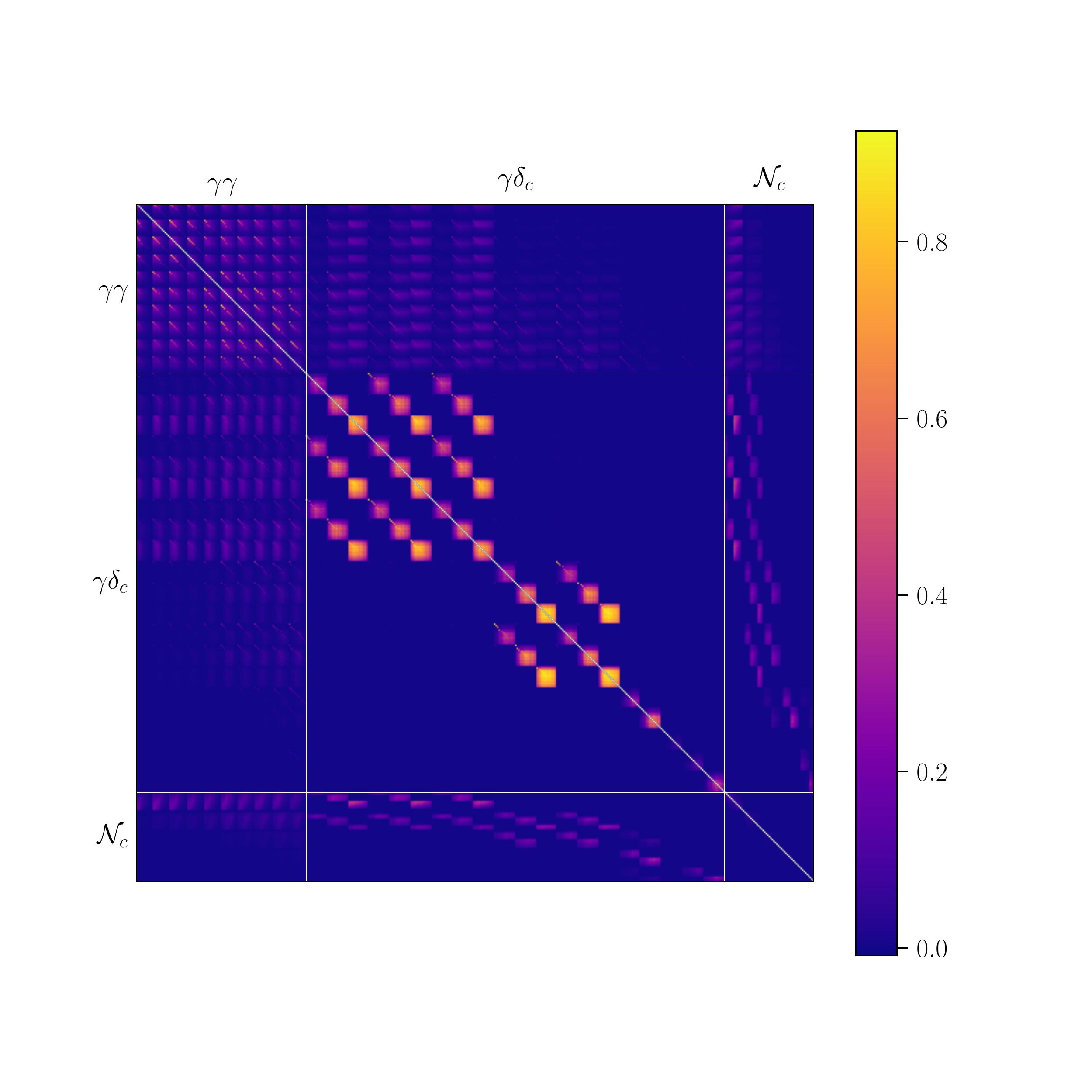}
 \caption{Joint correlation matrix of tSZ cluster number counts, cosmic shear and the cross-correlation between cluster overdensity and cosmic shear obtained in this analysis.}
\label{fig:covariance-fid}
\end{center}
\end{figure*}

We forecast constraints on cosmological and mass calibration parameters from a joint analysis of cluster number counts, cosmic shear and cluster lensing power spectra, assuming a Gaussian likelihood given by 
\begin{equation}
\mathscr{L}(\mathbf{D}^{\mathrm{obs}} \vert \theta) = \frac{1}{[(2\pi)^{d}\det{\mathbf{C}}]^{\sfrac{1}{2}}} e^{-\frac{1}{2}(\mathbf{D}^{\mathrm{obs}}-\mathbf{D}^{\mathrm{theor}})^{\mathrm{T}}\mathbf{C}^{-1}(\mathbf{D}^{\mathrm{obs}}-\mathbf{D}^{\mathrm{theor}})},
\label{eq:likelihood}
\end{equation}
where $\mathbf{C}$ denotes the non-Gaussian covariance matrix, computed as outlined in Sec.~\ref{sec:covariance}\footnote{We note that when computing the inverse covariance matrix, we first invert the correlation matrix and then transform back to the inverse covariance matrix. This avoids numerical instabilities due to the large dynamic range in the covariance matrix elements.} . Furthermore, $\mathbf{D}^{\mathrm{obs}}$ is the observed data vector, given by 
\begin{multline}
\mathbf{D}^{\mathrm{obs}} = (C^{\gamma_{i1} \gamma_{j1}}_{\ell}, \cdots, C^{\gamma_{in} \gamma_{jn}}_{\ell}, \; C_{\ell}^{\gamma_{i1}  \delta^{k1}_{\mathrm{cl}, \alpha 1}}, \cdots,  C_{\ell}^{\gamma_{im}  \delta^{km}_{\mathrm{cl}, \alpha m}}, \\
\mathcal{N}^{l1}_{\mathrm{cl}, \beta 1}, \cdots, \mathcal{N}^{lo}_{\mathrm{cl}, \beta o})_{\mathrm{obs}},
\label{eq:psvector}
\end{multline} 
and $\mathbf{D}^{\mathrm{theor}}$ denotes the corresponding theoretical prediction. The correlation matrix obtained in our analysis for the experimental specifications given in Sec.~\ref{sec:lsstxso} is shown in Fig.~\ref{fig:covariance-fid}\footnote{The correlation matrix $\mathsf{Corr}$ is obtained from the covariance matrix $\mathbf{C}$ as $\mathsf{Corr}_{ij} = \sfrac{\mathbf{C}_{ij}}{\sqrt{\mathbf{C}_{ii} \mathbf{C}_{jj}}}$.}. The full matrix has dimensions $(n, n) = (519, 519)$  and consists of $130$ $C^{\gamma \gamma}_{\ell}$ measurements, $320$ $C_{\ell}^{\gamma  \delta_{\mathrm{cl}}}$ measurements and $69$ $\mathcal{N}_{\mathrm{cl}}$ measurements. As can be seen, the different probes are significantly correlated and the importance of non-Gaussian contributions to the covariance, which give rise to the off-diagonal elements, increase with angular multipole $\ell$ and tSZ amplitude $Y^{\mathrm{obs}}_{500}$.

Traditionally, tSZ cluster mass calibration has been performed in a two step process: in a first step, cosmic shear, CMB lensing or X-ray measurements are used to derive prior constraints on cluster masses or mass calibration. In a second step, these prior constraints are folded into the cluster number counts likelihood to derive constraints on cosmological and mass calibration parameters. A number of different approaches exist in the literature (see e.g. \cite{Sehgal:2011, Bocquet:2015, Haan:2016, Alonso:2016, Louis:2017}), which vary in the data used to derive priors on mass calibration and their derivation. In order to further validate the mass calibration method proposed in this work, we compare its forecasted constraints to those obtained in such a stacking analysis. For the stacked cluster number counts likelihood, we closely follow the approach outlined in Ref.~\cite{Madhavacheril:2017}: we compute uncertainties on inferred weak lensing masses assuming measurements of the real-space cluster lensing signal for all clusters in the sample. These constraints are used to derive cluster number counts binned in redshift $z$, tSZ signal-to-noise $q$ and weak lensing mass $M_{\mathrm{WL}}$. The measurements are finally used to compute constraints on cosmological and mass calibration parameters assuming Poisson noise (i.e. neglecting the non-Gaussian covariance discussed above\footnote{We have made this choice in order to maintain consistency with the original analysis in Ref.~\cite{Madhavacheril:2017}}).

\section{Forecasting methods}\label{sec:forecasts}

We use a Fisher matrix formalism to forecast constraints on cosmological and mass calibration parameters for both methods outlined above. The Fisher matrix allows for propagation of experimental uncertainties to uncertainties on model parameters. Under the assumption that the dependence of the data covariance matrix on the parameters of interest $\theta_{\alpha}$ can be neglected, the Fisher matrix for a given experiment, measuring a data vector $\mathbf{D}$, is given by (see e.g. \cite{Fisher:1935, Kendall:1979, Tegmark:1997})
\begin{equation}
F_{\alpha \beta} = \frac{\partial \mathbf{D}}{\partial \theta_{\alpha}}\mathbf{C}^{-1}\frac{\partial \mathbf{D}}{\partial \theta_{\beta}}.
\end{equation}
The Cram\'er-Rao bound states that the uncertainty on $\theta_{\alpha}$, marginalized over all other $\theta_{\beta}$ satisfies
\begin{equation}
\Delta \theta_{\alpha} \geq \sqrt{(F^{-1})_{\alpha \alpha}}.
\end{equation}

Computing the Fisher matrix requires the assumption of a fiducial model. In this work, we choose cosmological parameter values close to those derived by the Planck Collaboration in their 2015 data release using only temperature data \cite{Planck:2016} (c.f. the first column of Tab. 4 in Ref.~\cite{Planck:2016}). The fiducial values assumed for all parameters are summarized in Tab.~\ref{tab:fiducial-model}.

We assess the potential of a combination of LSST and SO to simultaneously constrain cosmology and mass calibration by mainly investigating its constraining power on the time evolution of the dark energy equation of state parameter $w(a)$, parametrized as $w(a) = w_{0} + (1 - a)w_{a}$ \cite{Chevallier:2001, Linder:2003}\footnote{We note, however, that we expect the methods presented here to be useful for constraining any cosmological parameter affecting late-time structure growth, such as e.g. the sum of neutrino masses, $\sum_{i} m_{\nu, i}$.}. We therefore focus on $w_{0}w_{a}$CDM and forecast constraints on the set of cosmological and systematics parameters given by $\boldsymbol{\theta} = \{H_{0}, \allowbreak \Omega_{b}h^{2}, \allowbreak \Omega_{c}h^{2}, \allowbreak A_{s}, \allowbreak n_{s}, \allowbreak w_{0}, \allowbreak w_{a}, \allowbreak Y_{*}, \allowbreak \sigma_{\log{Y_{0}}}, \allowbreak \alpha_{\sigma}, \allowbreak \gamma_{\sigma}, \allowbreak \alpha_{Y}, \allowbreak \beta_{Y}, \allowbreak \gamma_{Y}, \allowbreak \Delta z_{i}, \allowbreak m_{i} \}$, $i \in [0, \cdots, 3]$, where $H_{0}$ is the Hubble parameter, $\Omega_{b}h^{2}$ is the physical baryon density today, $\Omega_{c}h^{2}$ is the physical cold dark matter density today, $n_{\mathrm{s}}$ denotes the scalar spectral index, $A_{s}$ is the primordial power spectrum amplitude at pivot wave vector $k_{0}=0.05$ Mpc$^{-1}$\footnote{We note that for consistency with Ref.~\cite{Madhavacheril:2017} we choose to parametrize the power spectrum amplitude in terms of $A_{s}$ instead of $\sigma_{8}$, which denotes the r.m.s. of linear matter fluctuations in spheres of comoving radius 8 $h^{-1}$ Mpc.} and $w_{0}, w_{a}$ parametrize the equation of state of dark energy. We compute derivatives of the observables with respect to these parameters numerically using a five-point stencil with step $\epsilon = 0.01\theta$, where $\theta$ denotes any parameter considered in our analysis\footnote{For parameters with fiducial value of zero, we set  $\epsilon = 0.01$.}. We test the stability of our results by varying the parameter $\epsilon$ and find our results to be largely insensitive to this choice. 

Unless stated otherwise, we combine our constraints with prior information from the Planck power spectrum following Ref.~\cite{Madhavacheril:2017}. Specifically, we include Planck temperature information from angular scales $2< \ell <30$ from the full Planck angular sky coverage ($f_{\mathrm{sky}}=  0.6$), temperature and polarization information from $30< \ell <100$ from the part of sky in which Planck and SO overlap ($f_{\mathrm{sky}}=  0.4$) and finally temperature and polarization information from $30< \ell <2500$ from the part of sky covered by Planck but not by SO ($f_{\mathrm{sky}}= 0.2$). Including the full Planck angular range and sky coverage, or the forecasted SO primary CMB information was found to not significantly impact forecasted constraints on $w_{0}$ and $w_{a}$ \cite{Ade:2019}, which are the primary focus of this work. We further follow Ref.~\cite{Krause:2017} and assume Gaussian priors on $\Delta z_{i}$ and $m_{i}$ with standard deviations $\sigma(\Delta z_{i})=0.002$ and $\sigma(m_{i})=0.004$ respectively. However, we do not assume any priors on the mass calibration parameters.

\begin{table*}
\caption{Summary of assumed fiducial model and parameters considered in the Fisher analysis.} \label{tab:fiducial-model}
\begin{center}
\begin{ruledtabular}
\begin{tabular}{cccc}
Parameter & Fiducial value & Prior & Description \\ \hline \Tstrut           
$H_{0}$ & 69. & Planck\footnote{See description in Sec.~\ref{sec:forecasts}.} & cosmology \\
$\Omega_{b}h^{2}$ & 0.02222 & Planck & cosmology \\
$\Omega_{c}h^{2}$ & 0.1197 &  Planck & cosmology \\
$A_{s}$ & $2.1955 \times 10^{-9}$ & Planck & cosmology \\
$n_{s}$ & 0.9655 & Planck & cosmology \\
$w_{0}$ & -1. & Planck & cosmology \\
$w_{a}$ & 0. & Planck & cosmology \\ \\     
$Y_{*}$ & $2.42 \times 10^{-10}$ & - & mean of $Y-M$ relation\footnote{See Eq.~\ref{eq:Y-M-mean}.} \\
$\alpha_{Y}$ & 1.79 & - & mean of $Y-M$ relation\\
$\beta_{Y}$ & 0. & - & mean of $Y-M$ relation\\
$\gamma_{Y}$ & 0. & - & mean of $Y-M$ relation \\
$\sigma_{\log{Y_{0}}}$ & 0.127 & - & scatter of $Y-M$ relation\footnote{See Eq.~\ref{eq:Y-M-scatter}.}  \\
$\alpha_{\sigma}$ & 0. & - & scatter of $Y-M$ relation\\
$\gamma_{\sigma}$ & 0. & - & scatter of $Y-M$ relation\\ \\
$\Delta z_{i}$ & 0. & $\mathcal{N}(\mu = 0, \sigma=0.002)$\footnote{Here, $\mathcal{N}$ denotes a 1-dimensional Gaussian distribution.} & photo-$z$ uncertainties \\ 
$m_{i}$ & 0. & $\mathcal{N}(\mu = 0, \sigma=0.004)$ & multiplicative shear bias \\
\end{tabular}
\end{ruledtabular}
\end{center}
\end{table*} 

\section{Results}\label{sec:results}

\begin{figure*}
\begin{center}
\includegraphics[width=0.95\textwidth]{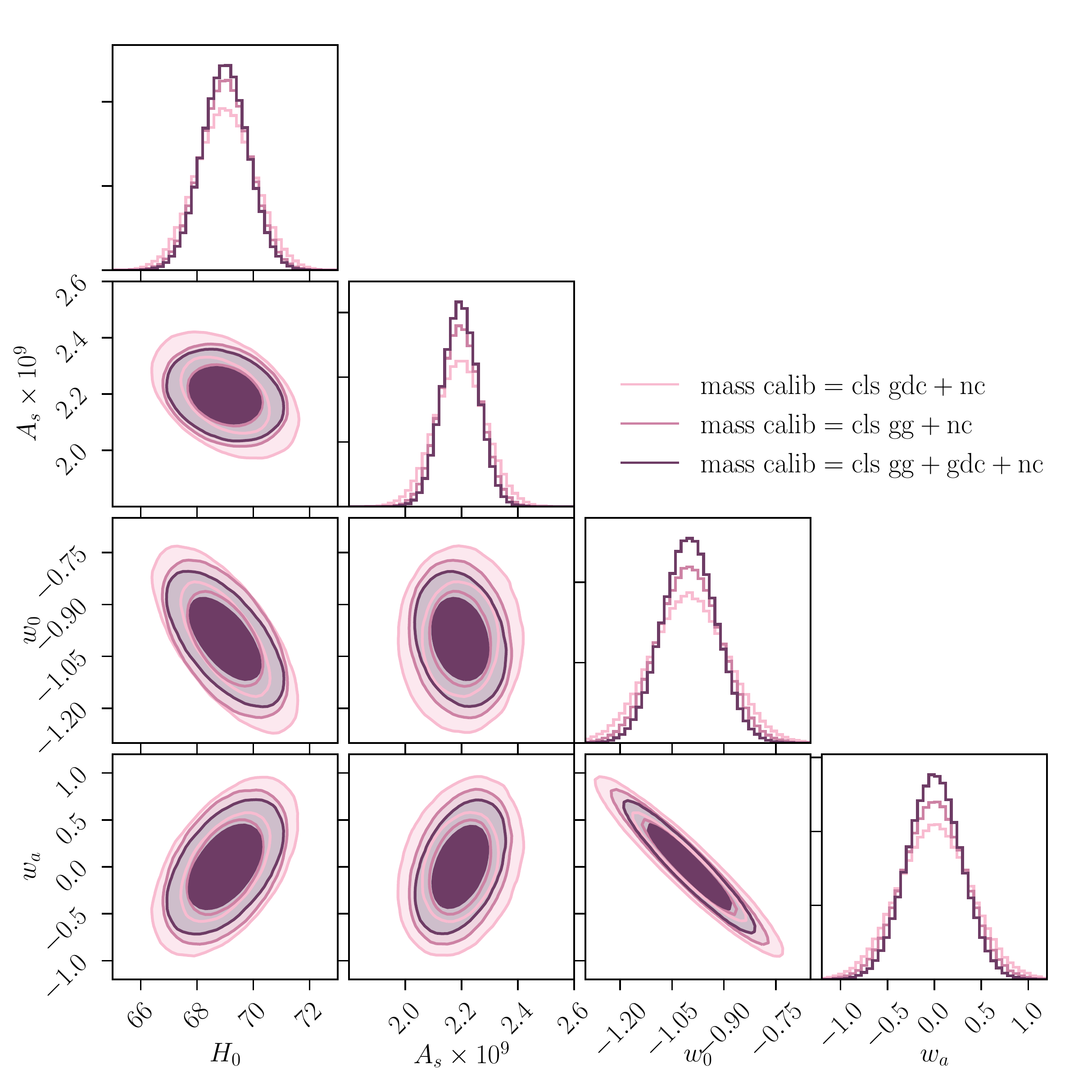}
 \caption{Forecasted constraints on a subset of cosmological parameters obtained in a joint analysis of LSST and SO for three different data splits. The constraints are marginalized over mass calibration and cosmic shear systematics parameters. The inner (outer) contour shows the $68 \%$ confidence limit (c.l.) ($95 \%$ c.l.).}
\label{fig:constraints-data-splits-cosmo-H0-As-w0-wa}
\end{center}
\end{figure*}

\begin{figure*}
\begin{center}
\includegraphics[width=0.95\textwidth]{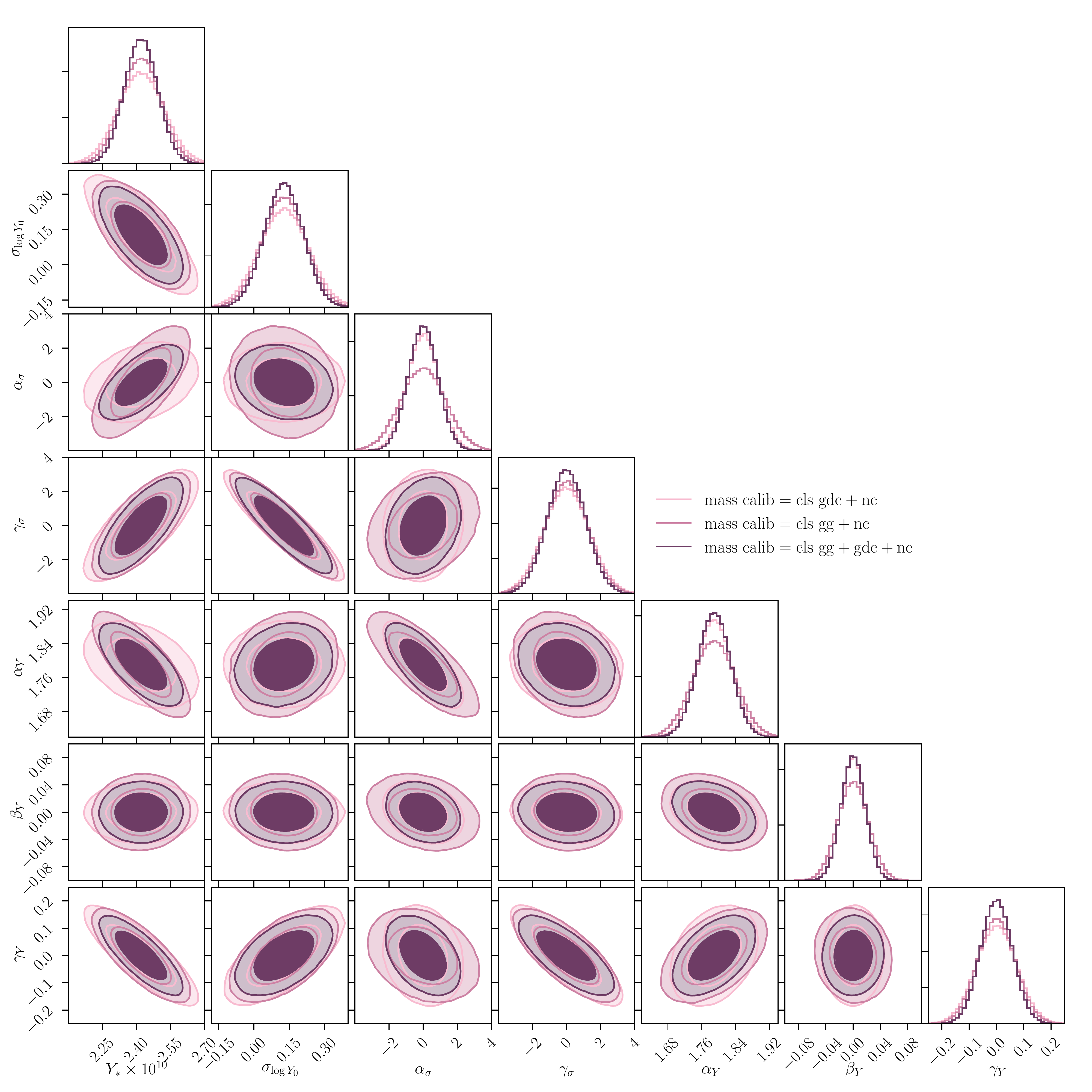}
 \caption{Forecasted constraints on mass calibration parameters obtained in a joint analysis of LSST and SO for three different data splits. The constraints are marginalized over cosmological and cosmic shear systematics parameters. The inner (outer) contour shows the $68 \%$ c.l. ($95 \%$ c.l.).}
\label{fig:constraints-data-splits-Y-M}
\end{center}
\end{figure*}

Fig.~\ref{fig:constraints-data-splits-cosmo-H0-As-w0-wa} shows our fiducial forecasted constraints on a subset of cosmological parameters\footnote{The full panel is shown in Fig.~\ref{fig:constraints-data-splits-cosmo} in the Appendix.} for a combination of LSST and SO, denoted $\texttt{gg+gdc+nc}$\footnote{Here, $\texttt{gg}$ denotes cosmic shear, $\texttt{gdc}$ denotes the cross-correlation between cluster overdensity and cosmic shear and finally $\texttt{nc}$ denotes cluster number counts.} in the figure. These constraints are obtained from a joint analysis of SO tSZ cluster number counts, LSST cosmic shear and the cross-correlation between cosmic shear and cluster overdensity, combined with prior information from Planck as described in Sec.~\ref{sec:forecasts}. The corresponding constraints on mass calibration parameters are shown in Fig. ~\ref{fig:constraints-data-splits-Y-M}. As can be seen, the combination of SO clusters with LSST cosmic shear has the potential to provide rather tight constraints on both cosmological and mass calibration parameters. As an example, the dark energy equation of state parameters $w_{0}$ and $w_{a}$ are constrained to a level of $\sim 8 \%$ and $\sigma(w_{a})\sim 0.3$, respectively. This constitutes an improvement in the Dark Energy Task Force (DETF) Figure of Merit \cite{Albrecht:2006} with respect to LSST cosmic shear alone of approximately a factor of two. In addition, we find that this combination provides tight constraints on $H_{0}$ and $A_{s}$, improving the uncertainties on the primordial power spectrum amplitude by a factor of two, again compared to LSST cosmic shear. These results also imply tighter constraints on $\sigma_{8}$, which is directly constrained by low-redshift large-scale structure observables. Comparing our fiducial constraints to those obtained from the Planck CMB prior alone, we find significant improvements in the constraints on $H_{0}$, $A_{s}$, $w_{0}$ and $w_{a}$, with the dark energy figure of merit increasing by a factor of approximately $1400$. Looking at the mass calibration parameters, we find a $\sim 3 \%$ constraint on the amplitude of the $Y-M$ relation, $Y_{*}$. Comparing this constraint to existing measurements is complicated by the fact that the respective analyses significantly differ in both methodology and constrained parameter set. We note however that this constraint constitutes a significant improvement compared to current constraints, which are at the level of $~ 17 \%$ (see e.g. Ref.~\cite{Bocquet:2019}). These results are especially remarkable, as the cosmological constraints are fully and self-consistently marginalized over uncertainties in the tSZ $Y-M$-relation and  cosmic shear measurement systematics and are derived accounting for the full non-Gaussian covariance between cluster number counts and the various cosmic shear observables. Similarly, the constraints on mass calibration shown in Fig.~\ref{fig:constraints-data-splits-Y-M} illustrate the constraining power of LSST and SO when self-consistently marginalizing over cosmic shear systematics.

In order to disentangle the contribution of separate probes to these constraints, we compute forecasted constraints for two subsets of our full data vector: in the first case, we combine only cosmic shear and cluster number counts (denoted $\texttt{gg+nc}$), and in the second case we combine cluster number counts and the cluster lensing power spectrum (denoted $\texttt{gdc+nc}$). The obtained constraints are shown in Figures \ref{fig:constraints-data-splits-cosmo-H0-As-w0-wa} and \ref{fig:constraints-data-splits-Y-M} alongside our fiducial ones. From these figures we see that the combination $\texttt{gg+nc}$ yields cosmological parameter constraints comparable to those obtained from our fiducial case, while leading to significantly weaker constraints on mass calibration. The combination $\texttt{gdc+nc}$ on the other hand, shows the opposite behavior, i.e. the cosmological constraints are weaker while the constraints on mass calibration are comparable to the fiducial case. These results suggest that adding cosmic shear to cluster number counts mainly affects the cosmological constraining power. Combining cluster lensing and number counts on the other hand, allows for precise mass calibration and breaks some of the degeneracies between cosmology and the $Y-M$ relation, inherent to cluster counts alone. 

It is interesting to ask which angular scales in $C_{\ell}^{\gamma \delta_{\mathrm{cl}}}$ contribute most to the constraints on the $Y-M$ relation. To this end, we forecast constraints for $\texttt{gdc+nc}$ restricting the angular multipole range for the cluster lensing cross-correlation to $\ell \leq 3000$ as compared to our fiducial case with $\ell \leq 9400$. Somewhat surprisingly, we find almost identical constraints on both cosmological and mass calibration parameters in both cases\footnote{As the constraints are almost indistinguishable, we do not show them in any of the figures. In addition, further reducing the multipole range to $\ell \leq 1000$ only leads to modest increases in parameter uncertainties.}. This suggests that the constraints on mass calibration are driven by the large and intermediate angular scales rather than the smallest scales considered in our analysis. As can be seen from Fig.~\ref{fig:observables}, these scales receive contributions from both the 1- and the 2-halo term of the power spectrum. For the intermediate redshift bin shown in Fig.~\ref{fig:observables}, the 2- to 1-halo transition occurs at $\ell \sim 300$, while for the highest redshift bin, it is pushed to $\ell \sim 600$. Our results thus suggest that the amplitude of $C_{\ell}^{\gamma \delta_{\mathrm{cl}}}$ on relatively large angular scales contains some information on mass calibration, as also seen in Ref.~\cite{Majumdar:2004}. The large-scale amplitude of the cluster lensing signal is predominantly determined by the cluster bias, which depends on mass, and is therefore sensitive to mass calibration parameters, thus allowing for constraining the mass-observable relation. This is different from traditional mass calibration methods, which solely focus on the 1-halo term and thus use information from smaller scales to constrain the $Y-M$ relation\footnote{A potential concern about using information from the large-scale amplitude of the cluster lensing signal for mass calibration is the uncertainty on cluster bias models. In order to test the robustness of our results to these uncertainties, we forecast constraints from $\texttt{gdc+nc}$ accounting for a $10 \%$ uncertainty in the amplitude of the cluster lensing power spectrum, finding only modest increases in parameter constraints.}. This complementarity therefore suggests an interesting way to test for systematics in mass calibration by comparing the results obtained with both methods. 

\begin{figure*}
\begin{center}
\includegraphics[width=0.95\textwidth]{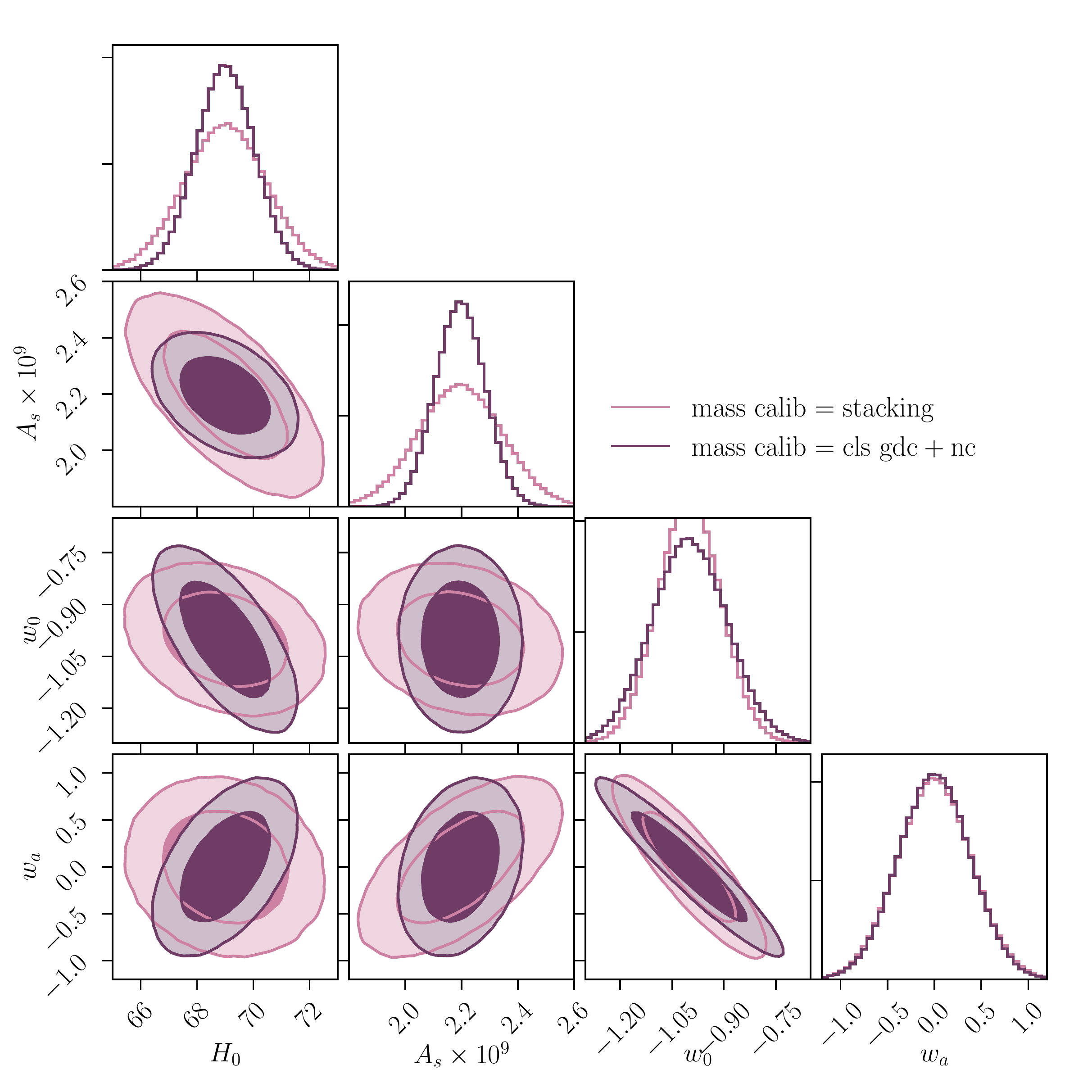}
 \caption{Comparison of the forecasted constraints on a subset of cosmological parameters obtained using the two methods outlined in Sec.~\ref{sec:mass-calib}. The constraints are marginalized over mass calibration and cosmic shear systematics parameters. The inner (outer) contour shows the $68 \%$ c.l. ($95 \%$ c.l.).}
\label{fig:constraints-cosmo-cls-vs-stacking-H0-As-w0-wa}
\end{center}
\end{figure*}

We further test the methodology presented in this analysis by comparing the obtained forecasted constraints to those obtained performing a traditional stacking analysis, as described in Sec.~\ref{sec:mass-calib}. As the stacking analysis does not contain cosmic shear information, we only perform this comparison for the $\texttt{gdc+nc}$ data split. We constrain the same parameter set and apply identical priors to both methods, except that for consistency with existing analyses we do not account for cosmic shear systematics when forecasting constraints from the stacking method. The resulting constraints are shown in Figures \ref{fig:constraints-cosmo-cls-vs-stacking-H0-As-w0-wa}\footnote{The full panel for the cosmological parameter constraints is shown in Fig.~\ref{fig:constraints-cosmo-cls-vs-stacking} in the Appendix.} and \ref{fig:constraints-Y-M-cls-vs-stacking}. As opposed to the constraints obtained from the stacking method, the constraints from the cross-correlation method are fully marginalized over cosmic shear systematic uncertainties and are derived taking into account the full non-Gaussian covariance between cluster counts and cosmic shear. From Figures \ref{fig:constraints-cosmo-cls-vs-stacking-H0-As-w0-wa} and \ref{fig:constraints-Y-M-cls-vs-stacking} we see that the cross-correlation method nevertheless yields significantly tighter constraints on cosmological parameters, especially $H_{0}$ and $A_{s}$, where we find a reduction in the $1\sigma$ uncertainties of approximately $30\%$ and $40\%$ respectively. For the mass calibration, we find the cross-correlation method to yield comparable or tighter constraints on the parameters entering the mean of the $Y-M$ relation (see Eq.~\ref{eq:Y-M-mean}), e.g. $\beta_{Y}$. In contrast however, the obtained constraints on the scatter in the $Y-M$ relation (see Eq.~\ref{eq:Y-M-scatter}) are weaker. From Fig.~\ref{fig:constraints-Y-M-cls-vs-stacking} we see that the larger uncertainties on these parameters are mainly driven by increased parameter degeneracies obtained for the cross-correlation method. This suggests that these differences are not due to the mass calibration method itself but rather due to the different treatment of cluster number counts in both analyses: while the stacking method allows for binning the cluster number counts in both $M_{\mathrm{WL}}$ and $Y^{\mathrm{obs}}_{500}$, the number counts in the cross-correlation method are only binned in $Y^{\mathrm{obs}}_{500}$. This lack of explicit mass information in the cluster number counts can lead to larger degeneracies and thus enhanced correlations between the different mass calibration parameters. Further confirmation comes from the fact that we find the derivatives of the stacked cluster number counts with respect to the parameters of $\sigma_{\log Y_{500}}(M, z)$ marginalized over $Y^{\mathrm{obs}}_{500}$ to be significantly larger than the derivatives obtained when marginalizing the cluster number counts over $M_{\mathrm{WL}}$. Another way of seeing this is that we find a loss of most of the constraining power on the scatter of the $Y-M$ relation when using the stacked cluster number counts marginalized over $M_{\mathrm{WL}}$. As discussed above, an additional reason for these differences might be the fact that the constraints derived using the cross-correlation method are fully marginalized over systematics in the cosmic shear and take into account the cross-correlation between cluster number counts and cosmic shear, in contrast to the stacking method.

Despite the somewhat weaker constraints on the scatter in the $Y-M$ relation, these results show that the cluster lensing power spectrum provides a promising alternative to traditional tSZ mass calibration methods, as it allows for both precise mass calibration and additionally provides cosmological information complementary to cluster number counts (as can be seen from the fact that the cosmological constraints from $\texttt{gdc+nc}$ are tighter than those obtained with the stacking method). 

\begin{figure*}
\begin{center}
\includegraphics[width=0.95\textwidth]{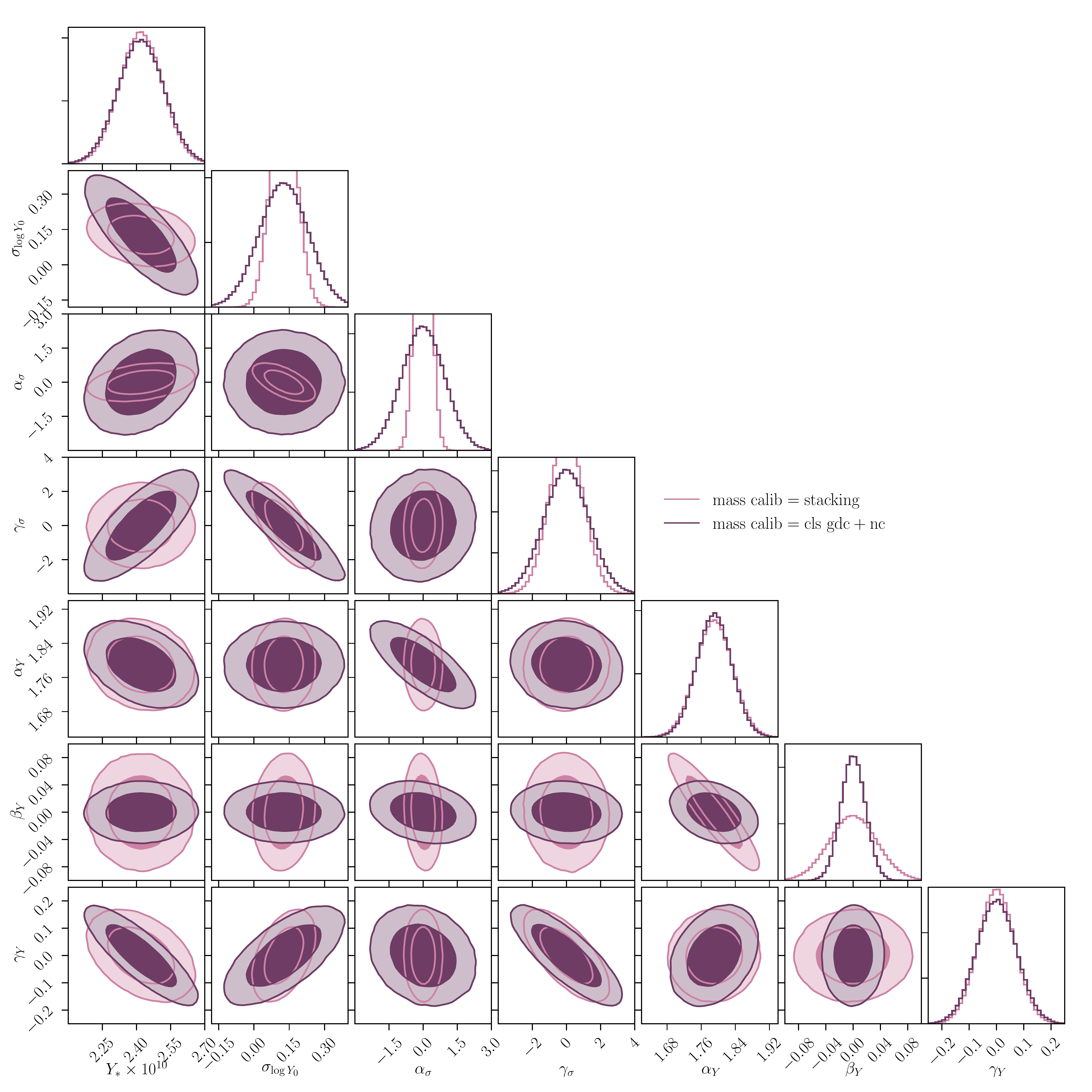}
 \caption{Comparison of the forecasted constraints on mass calibration parameters obtained using the two methods outlined in Sec.~\ref{sec:mass-calib}. The constraints are marginalized over cosmological and cosmic shear systematics parameters. The inner (outer) contour shows the $68 \%$ c.l. ($95 \%$ c.l.).}
\label{fig:constraints-Y-M-cls-vs-stacking}
\end{center}
\end{figure*}

\section{Summary and conclusions}\label{sec:conclusions}

In this work we present a novel method for joint cosmological parameter inference and cluster mass calibration from a combination of weak lensing measurements and thermal Sunyaev-Zel'dovich cluster abundances. We focus on a combination of cluster number counts, angular cosmic shear power spectra and the angular cross-correlation between cluster overdensity and cosmic shear, which acts as the main cluster mass calibrator in our analysis. Using a halo model approach, we derive and compute theoretical estimates for all observables as well as their full non-Gaussian covariance. We then forecast constraints for a joint analysis of LSST and SO on both cosmological and mass calibration parameters in a Fisher analysis, fully marginalizing over systematic uncertainties in cosmic shear measurements. Our results show that the method presented here yields competitive constraints on both cosmological and mass calibration parameters. Furthermore, we find most of the mass calibration information to be contained in the large and intermediate angular scales of the cross correlation between cosmic shear and cluster overdensity.   

We then compare our constraints to those obtained in a more traditional stacked cluster weak lensing analysis. Generally, we find the method presented here to yield tighter constraints on cosmological parameters and comparable or tighter constraints on the mean of the mass-observable relation. However, we find the scatter in the mass-observable relation to be more strongly constrained with the traditional method. We attribute this not to the mass calibration method itself but rather to different treatments of cluster number counts in both methods: the traditional methods allow for binning the cluster number counts in mass $M_{\mathrm{WL}}$ and tSZ amplitude $Y^{\mathrm{obs}}_{500}$ while the cluster counts in the method presented here are solely binned in $Y^{\mathrm{obs}}_{500}$. The additional mass binning in traditional methods allows to break degeneracies between the parameters of the mass-observable relation and therefore leads to tighter constraints on its scatter. 

Therefore, our analysis shows that the cross-correlation between cluster overdensity and cosmic shear provides a promising alternative to traditional mass calibration methods, offering several advantages compared to traditional approaches. First of all, the constraints derived using the method presented here are fully and consistently marginalized over cosmic shear measurement systematics and are derived taking into account the full non-Gaussian covariance between cluster counts and cosmic shear. Secondly, computing the cross-correlation between cosmic shear and cluster overdensity amounts to performing a statistical mass calibration. In contrast, traditional mass calibration methods require measuring the cluster lensing signal for each cluster in the sample, which might become prohibitively expensive for future surveys. Finally, the joint cluster count and cosmic shear likelihood derived in this work can be readily combined with other probes of the large-scale structure, such as galaxy clustering.

We envisage several possible extensions of the present work. On the one hand it will be interesting to test the method presented here by applying it to combinations of current CMB and large-scale structure surveys, such as ACT, SPT or DES. Due to the lower signal-to-noise in these data, as compared to LSST and SO, we however expect to constrain only a subset of the parameters considered in this work, especially those entering the mass calibration. Furthermore, applying this method to data will necessitate the inclusion of additional systematics, such as baryon feedback effects on the matter power spectrum (see e.g. Refs.~\cite{Rudd:2008, vanDaalen:2011}). On the theoretical side, we aim to investigate the potential of the cross-correlation method to constrain non-parametric mass-observable relations, which would remove the need of assuming uncertain functional forms for both the mean and scatter of the $Y-M$ relation. 

The analysis presented in this work shows that the cross-correlation method provides a promising and self-consistent way for jointly analyzing thermal Sunyaev-Zel'dovich cluster counts and cosmic shear. This bodes well for paving the way for multi-probe analyses including tSZ cluster number counts and harnessing the full potential of galaxy clusters as a precision cosmological probe.

\begin{acknowledgments}
AN would especially like to thank An\v{z}e Slosar and David Alonso: An\v{z}e Slosar for pulling one of his many Eastern European tricks when AN was stuck on this project and for comments on an earlier version of this manuscript. David Alonso for encouragement and many helpful discussions and comments on an earlier version of this manuscript. We would further like to thank Mat Madhavacheril and Nick Battaglia for many useful discussions regarding stacked weak lensing mass calibration and for help with using their code \texttt{szar}\footnote{\url{https://github.com/nbatta/szar}.}. We would also like to thank Mat Madhavacheril and Colin Hill for comments on an earlier version of this manuscript. Finally, we would like to thank Elisabeth Krause for helpful discussions regarding covariance matrices and Emmanuel Schaan for helpful discussions and for code comparison.

JD and AN acknowledge support from National Science Foundation Grant No. 1814971. The Flatiron Institute is supported by the Simons Foundation.

This is not an official SO collaboration paper.

The color palettes employed in this work are taken from $\tt{http://colorpalettes.net}$. The contour plots have been created using $\tt{corner.py}$ \cite{ForemanMackey:2016}.

\end{acknowledgments}

\appendix

\section{Transforming between mass definitions}\label{ap:sec:mass-trans} 

Throughout this work, we need to transform between different mass definitions. The total halo mass enclosed within a radius $R$ for an NFW density profile is given by
\begin{equation}
\begin{aligned}
M(< R) =& \; 4 \pi \int_{0}^{R} \mathrm{d}r \; r^{2} \rho_{\mathrm{NFW}}(r) = \\&4 \pi \rho_{0} r_{s}^{3}\left[\log\left(1 + \frac{R}{r_{s}}\right) - \frac{\sfrac{R}{r_{s}}}{1+\sfrac{R}{r_{s}}} \right],
\end{aligned}
\end{equation} 
where $r_{s}$ denotes the scale radius and $\rho_{0}$ the characteristic density of a given halo.
In the case in which $R = R_{\Delta}$, we obtain using $c_{\Delta} = \sfrac{R_{\Delta}}{r_{s}}$
\begin{equation}
M(< R_{\Delta}) = 4 \pi \rho_{0} r_{s}^{3}\left[\log\left(1 + c_{\Delta}\right) - \frac{c_{\Delta}}{1+c_{\Delta}} \right].
\end{equation} 
Therefore we obtain a relation between halo masses defined using different overdensity criteria $\Delta$ as
\begin{equation}
\frac{M(< R_{\Delta})}{M(< R_{\Delta'})} = \frac{\log\left(1 + c_{\Delta}\right) - \frac{c_{\Delta}}{1+c_{\Delta}}}{\log\left(1 + c_{\Delta'}\right) - \frac{c_{\Delta'}}{1+c_{\Delta'}}}.
\label{eq:mass-def-trans}
\end{equation} 
The above equation is an implicit function of $M_{\Delta'} \equiv M(< R_{\Delta'})$. In this work, we convert between $M$ and $M_{500}$ by iteratively solving Eq.~\ref{eq:mass-def-trans}.

\section{Implementation details}\label{ap:sec:implementation}

\subsection{Cluster counts binning scheme}\label{ap:sec:implementation.counts}

We first divide the distribution of galaxy clusters into five redshift bins with bin edges $z_{i} \in [0., 0.25, 0.5, 0.75, 1., 1.5]$. As discussed in Sec.~\ref{sec:lsstxso}, we employ different tSZ amplitude bins for each redshift bin, in order to ensure at least one cluster per bin in all cases. For the first redshift bin, we consider 15 logarithmically-spaced bins between $Y^{\mathrm{obs}}_{500, \mathrm{min}} = 8.6\times 10^{-12}$ and $Y^{\mathrm{obs}}_{500, \mathrm{max}} = 3.9\times 10^{-9}$. For the second redshift bin, we consider 14 logarithmically-spaced bins between $Y^{\mathrm{obs}}_{500, \mathrm{min}} = 4.3\times 10^{-12}$ and $Y^{\mathrm{obs}}_{500, \mathrm{max}} = 5.1\times 10^{-10}$. For the third redshift bin, we consider 15 logarithmically-spaced bins between $Y^{\mathrm{obs}}_{500, \mathrm{min}} = 3.1\times 10^{-12}$ and $Y^{\mathrm{obs}}_{500, \mathrm{max}} = 1.8\times 10^{-10}$. For the fourth redshift bin, we consider 13 logarithmically-spaced bins between $Y^{\mathrm{obs}}_{500, \mathrm{min}} = 3.1\times 10^{-12}$ and $Y^{\mathrm{obs}}_{500, \mathrm{max}} = 1.1\times 10^{-10}$. And finally for the fifth redshift bin, we consider 12 logarithmically-spaced bins between $Y^{\mathrm{obs}}_{500, \mathrm{min}} = 2.5\times 10^{-12}$ and $Y^{\mathrm{obs}}_{500, \mathrm{max}} = 6.6\times 10^{-11}$.

\subsection{Cluster lensing power spectrum binning scheme}\label{ap:sec:implementation.clust-lens}

We compute the cross-correlation between cosmic shear and cluster overdensity in four redshift bins with bin edges $z_{i} \in [0., 0.35, 0.7, 1.05, 1.41]$. We further subdivide these redshift bins into four logarithmically-spaced tSZ amplitude bins between $Y^{\mathrm{obs}}_{500, \mathrm{min}} = 4\times 10^{-13}$ and $Y^{\mathrm{obs}}_{500, \mathrm{max}} = 1.4\times 10^{-9}$. Requiring that each bin contain at least a single cluster removes five of these bins, which leaves us with 11 out of our original 16 tSZ amplitude and redshift bins.

\begin{figure*}
\begin{center}
\includegraphics[width=0.95\textwidth]{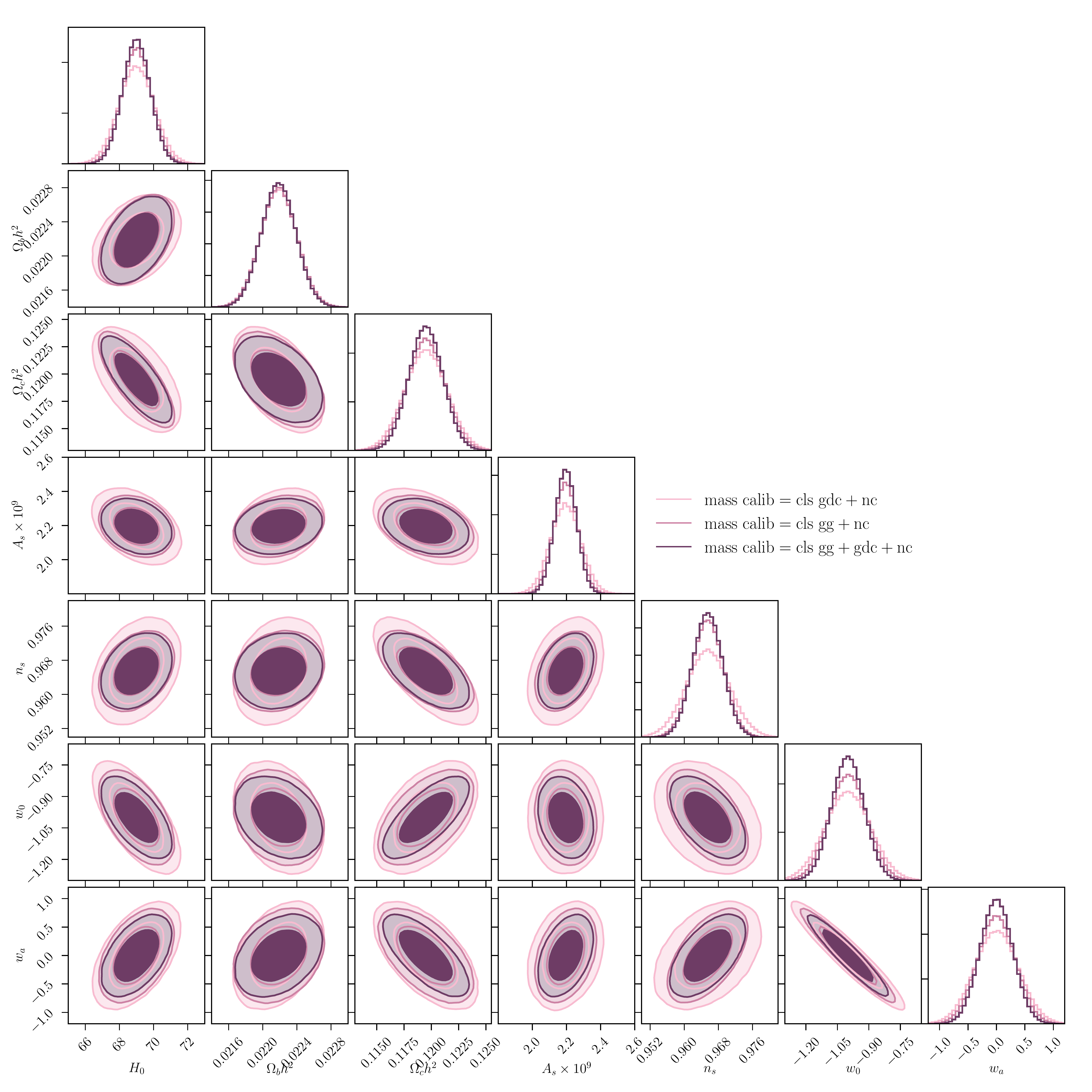}
 \caption{Forecasted constraints on cosmological parameters obtained in a joint analysis of LSST and SO for three different data splits. The constraints are marginalized over mass calibration and cosmic shear systematics parameters. The inner (outer) contour shows the $68 \%$ c.l. ($95 \%$ c.l.).}
\label{fig:constraints-data-splits-cosmo}
\end{center}
\end{figure*}

\begin{figure*}
\begin{center}
\includegraphics[width=0.95\textwidth]{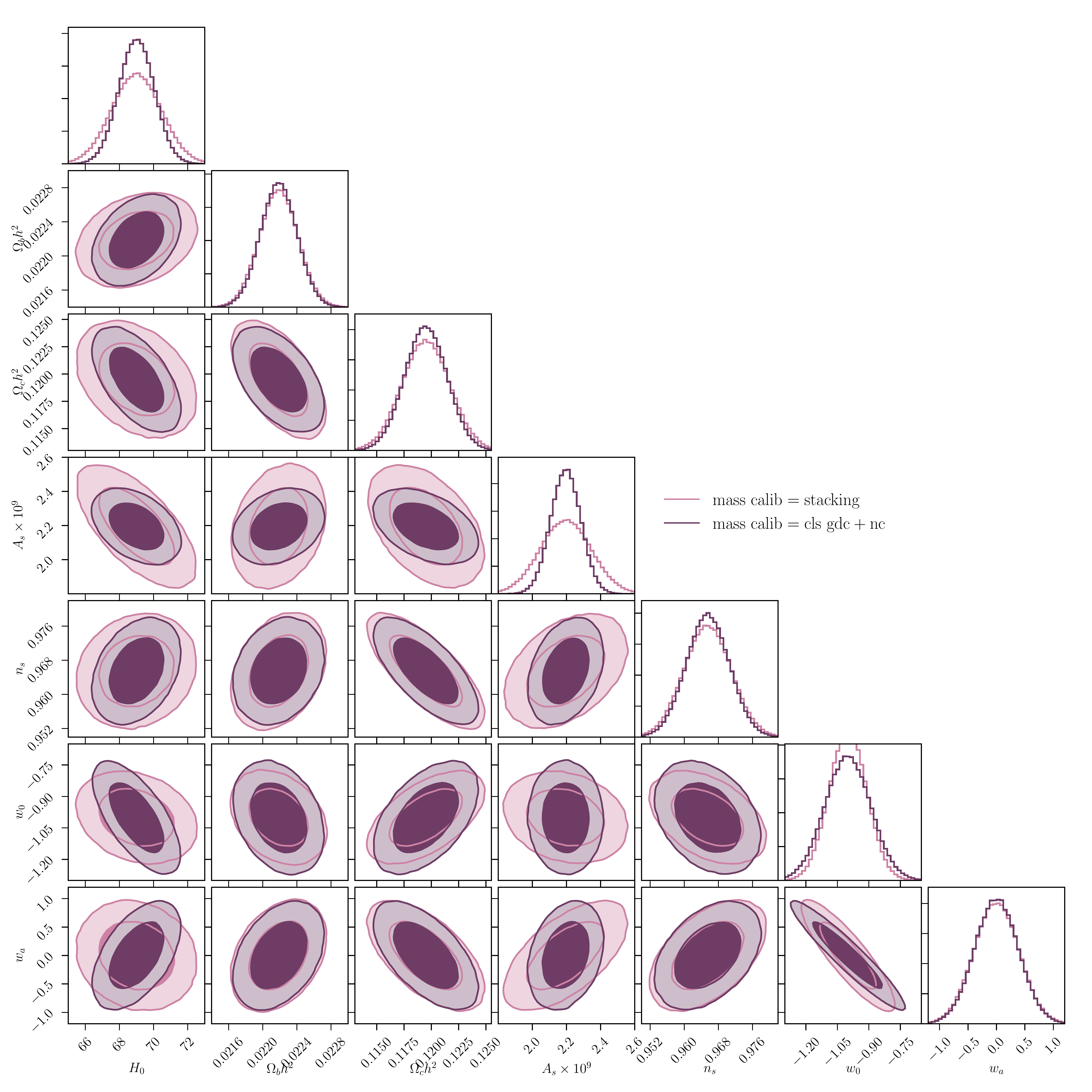}
 \caption{Comparison of the forecasted constraints on cosmological parameters obtained using the two methods outlined in Sec.~\ref{sec:mass-calib}. The constraints are marginalized over mass calibration and cosmic shear systematics parameters. The inner (outer) contour shows the $68 \%$ c.l. ($95 \%$ c.l.).}
\label{fig:constraints-cosmo-cls-vs-stacking}
\end{center}
\end{figure*}

\bibliography{main_text_incl_figs}% Produces the bibliography via BibTeX.

\end{document}